\documentclass[
	aps,
	prstab,
	a4paper,
	amsmath,
	amssymb,
	floatfix,
	showpacs,
	reprint,
	]{revtex4-1}
	
\usepackage[utf8]{inputenc}

\usepackage{graphicx}\graphicspath{./graph/}%
\usepackage{dcolumn}% Align table columns on decimal point
\usepackage{bm}% bold math
\usepackage{color}
\usepackage{verbatim}
\usepackage{amssymb}
\usepackage{amsfonts}
\usepackage{latexsym}
\usepackage{epstopdf}
\usepackage[alsoload=hep]{siunitx}
\usepackage{makecell}
\usepackage{multirow}
\usepackage[caption=false]{subfig}
\usepackage{lineno}
\usepackage[inline,shortlabels]{enumitem}

\begin{document}
%\preprint{Tech. Note ALSU-AP-TN-2019-XX}

\title{The Three-Dipole Kicker Injection Scheme for the ALS-U Accumulator Ring}
\author{M. Ehrlichman, T. Hellert, S.C. Leemann, G. Penn,  C. Steier, C. Sun,  M. Venturini, and D. Wang}
\affiliation{Lawrence Berkeley National Laboratory, Berkeley, CA}

\date{\today}
%\date{February 2019}

\begin{abstract}
The ALS-U light source will implement  on-axis single-train swap-out injection employing an accumulator between the 
booster and storage rings.  The accumulator ring design is a twelve period triple-bend achromat that will 
be installed along the inner circumference of the storage-ring tunnel. A non-conventional injection scheme will be utilized for top-off off-axis injection from the booster into the accumulator ring meant to accommodate a large $\sim 300$~nm emittance beam into a vacuum-chamber with a limiting horizontal aperture radius as small as $8$ mm.  The scheme 
incorporates three dipole kickers distributed over three sectors,  with two kickers perturbing  the 
stored beam and the third affecting both the stored and the injected beam trajectories. This paper describes 
this ``3DK'' injection scheme and how it fits the accumulator ring's particular requirements.
We describe the design and optimization process, and how we evaluated its fitness 
as a solution for booster-to-accumulator ring injection.
\end{abstract}    
    
\maketitle

\section{Introduction}
The ALS-U is the Advanced Light Source upgrade to a 4th generation diffraction-limited soft x-ray light source \cite{Steier:IPAC2018-THPMF036}. Achieving the  ALS-U high brightness goal requires strong focusing-magnet gradients; the strong gradients necessitate strong chromatic sextupoles and these will   shrink the dynamic aperture of the storage ring (SR) to about $0.5$~mm radius once lattice imperfections are taken into account.

%To inject into such a small dynamic aperture, ALS-U will implement full-train swap-out injection utilizing an  accumulator ring (AR) housed along the inner wall of the SR tunnel, see Fig.~\ref{fig:machine} for an overview of  the ALS-U accelerator complex. The AR has two functions, one as a damping  ring  and the other as a beam-charge recycler in between swap-outs.  The SR  average current is $500$ mA, distributed evenly over a 284-bunch beam consisting of  eleven trains of $26$ (or $25$) bunches each.  The AR is designed to carry a single  train at a time, which is swapped with one of the SR trains once every about half a minute. The design emittance of the AR is $2$ nm-rad, much smaller than the approximately $300$ nm-rad emittance of  the bunches  delivered by the booster.   The $2$ nm-rad beam is required to inject into the storage ring with near-$100$\% efficiency and sufficient margin.  

\begin{figure*}[ht]
	\centering
 	\includegraphics*[width=\linewidth]{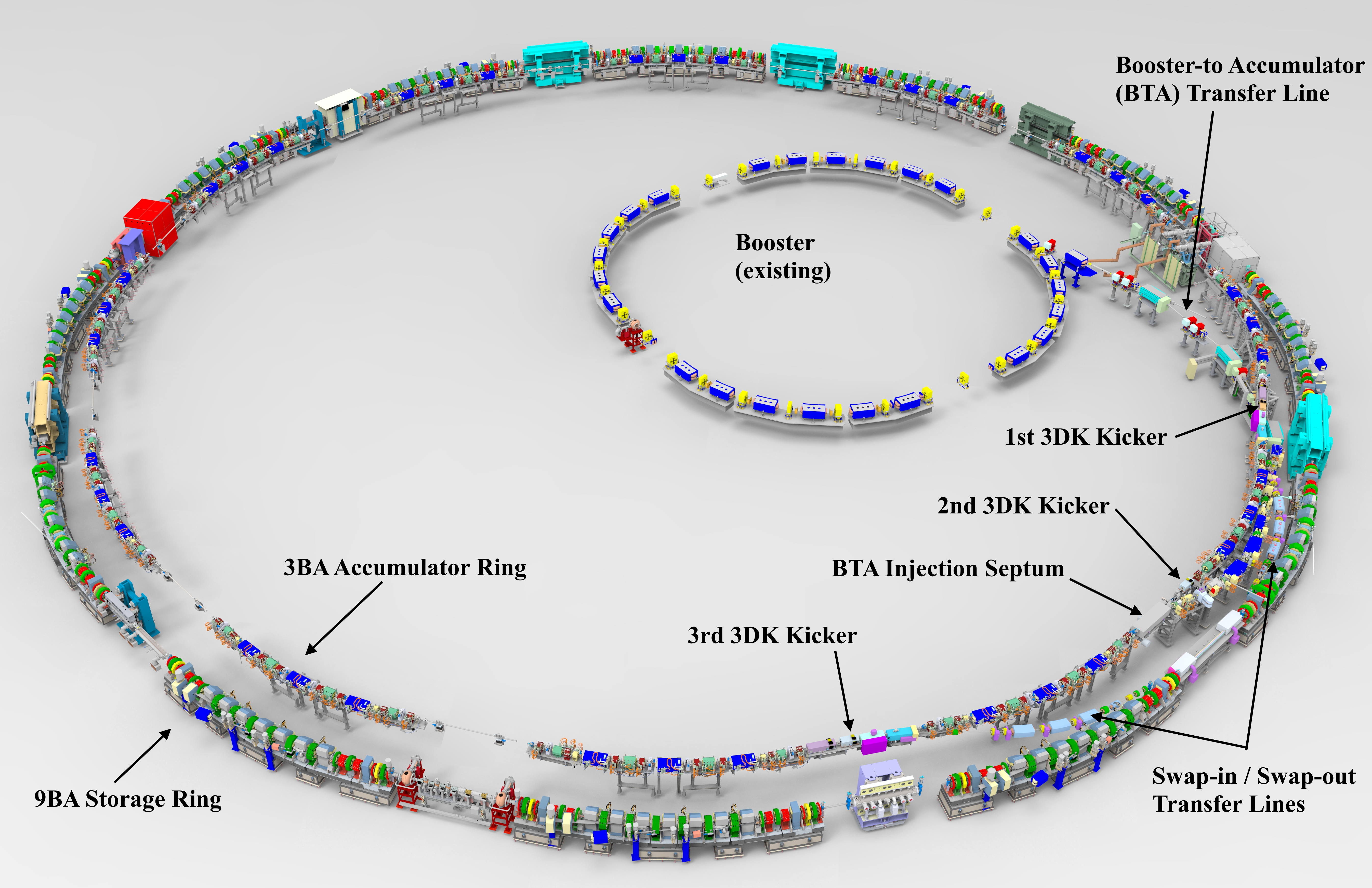}
	\caption{\label{fig:machine} The Advanced Light Source Upgrade (ALS-U) complex.  
	The $3$ dipole kickers and injection septum of the 3DK booster-to-accumulator
	injection scheme are highlighted.
	Every \SI{1.4}{\s} up to four bunches are generated and accelerated to \SI{2}{\GeV}
	through the existing linac and booster and then injected through the BTA into the
	new AR to fill or top-off bunches in a single 26- (or 25-) bunch train.  Every $\sim$\SI{30}{\s} this train is swapped with one of the eleven trains circulating in the	Storage Ring (SR).}
\end{figure*}

To inject into such a small dynamic aperture, ALS-U will implement on-axis single-train swap-out injection utilizing an  accumulator ring (AR) housed along the inner wall of the SR tunnel. See Fig.~\ref{fig:machine} for an overview of  the ALS-U accelerator complex. A small $2$ nm-rad natural emittance,  much smaller than the approximately $300$ nm-rad emittance of the beam delivered by  the existing booster,  is required to inject into the storage ring with near-$100$\% efficiency and sufficient margin. In addition to its function as a damping ring, the AR  is intended to act as a beam-charge recycler in between swap-outs.  The SR  average current is $500$ mA, distributed evenly over a 284-bunch beam consisting of  eleven trains of $26$ (or $25$) bunches each. The AR is designed to carry a single  train at a time; this is swapped with one of the SR trains once every about half a minute and replenished with top-off injection from the booster before the next swap-out.  
\begin{table}
\centering
\caption{\label{tab:AR} Parameter list of the ALS-U Accumulator Ring and existing ALS.}
\begin{ruledtabular}
\begin{tabular}{l|r|r}
& AR & ALS\footnote{Emittance and energy loss per turn reported for the ALS are for the lattice without superbends 
and without insertion devices.} \\
\hline
Beam energy                             & \SI{2.0}{\giga\electronvolt} & \SI{1.9}{\giga\electronvolt} \\
Circumference                           & \SI{182.122}{\meter} & \SI{196.8}{\meter} \\
Tune   $x/y$                                 & $16.221$/$8.328$ & $16.165$/$9.25$\\
Natural chrom.  $x/y$                        & $-43$/$-36$ & $-46.8$/$-39.6$ \\
Mom. compaction                         & \SI{1.04e-3}{} & \SI{0.9e-3}{} \\
Emittance                               & \SI{1.8}{\nano\meter} & \SI{1.8}{\nano\meter} \\
Dispersion in straight                  & \SI{11.6}{\centi\meter} & \SI{15.0}{\centi\meter} \\
Charge per bunch                        & \SI{1.15}{\nano\coulomb} & \SI{1.1}{\nano\coulomb} \\
Energy spread                           & \SI{8.5e-4}{} & \SI{9.6e-4}{} \\
Energy loss per turn                    & \SI{269}{keV}  & \SI{228}{keV} \\
Damping time  $x/y/z$                   & $6.2$/$8.5$/$5.2$ \SI{}{\milli\s} & $7.7$/$8.9$/$5.0$ \SI{}{\milli\s} \\
Harmonic number                         & \SI{304}{} & \SI{328}{} \\
Main RF frequency,                      & \SI{500.417}{MHz} & \SI{499.642}{MHz}\\
Main RF voltage,                        & \SI{1.0}{MV} & \SI{1.2}{MV} \\
Synchrotron tune                        & \SI{4.9e-3}{} & \SI{5.4e-3}{} \\
\end{tabular}
\end{ruledtabular}
\end{table}
%  \footnote{Energy loss per turn in the ALS includes superbends, which are not present in the AR.}

The AR design has to fulfill two competing demands on the vacuum-chamber aperture: it should be wide enough to accept the large emittance beam from the booster, with a goal of $\gtrsim 95\%$ injection efficiency,  but  as narrow as possible to minimize the magnets' aperture and thus their weight and volume 
so that the AR can fit  in the same tunnel as the SR.   Consideration of these demands has guided  the choice 
of injection scheme while  recognizing that unlike light sources the AR can tolerate significant injection transients. Fully exploiting the latitude offered by the latter observation, 
we have developed a simple and nearly 100\% efficient injection scheme, coined ``3DK'',  that utilizes three 
dipole kickers distributed over three sectors in combination with a pair of thick and thin  pulsed septa.

In short, the first two dipole kickers (``prekickers'') kick only the stored beam and place it on a 
trajectory that partly compensates the main injection kick placed further downstream, which kicks both the stored and 
injected particles onto stable (but not closed)  trajectories. The features of this scheme include
\begin{enumerate*}[1) ]
\item the reduced straight length is accommodated by distributing the kickers among $3$ straights,
\item near-100\% injection efficiency,
\item a reduced septum aperture that eases vacuum and magnet engineering in the injection straight,
\item flexibility to trade between residual oscillations of the stored and injected beams, and 
\item reliance on only conventional power supply and kicker technology.
\end{enumerate*}

An important aspect of our study is the demonstration that potential drawbacks resulting from the combination of 
relatively large stored-beam oscillations and collective effects are manageable.

Other schemes that were considered  and ultimately discarded included a non-linear kicker (NLK), a 
conventional $4$-kicker  single-straight closed-bump injection scheme, as well as a two-dipole kicker variant of the 3DK schemes.  These will be briefly discussed in this paper as well.

Section \ref{sec:ALS-U-AR} is a brief introduction to the ALS-U AR.  The layout and design of the 3DK 
injection scheme is described in Sec.~\ref{sec:3DK}. For completeness  Sec.~\ref{sec:BTA} briefly describes the design of the booster-to-AR transfer line.  The robustness of the
injection scheme against lattice imperfections is discussed in Sec.~\ref{sec:imperfections}.  Impedance and collective
effects, as well as multi-bunch feedback issues are addressed in Sec.~\ref{sec:TFB}.  Alternatives to the
3DK injection scheme are discussed in the appendix.

\section{ALS-U Accumulator Ring}\label{sec:ALS-U-AR}

\begin{figure}[ht]
	\centering
	\includegraphics[width=0.48\textwidth]{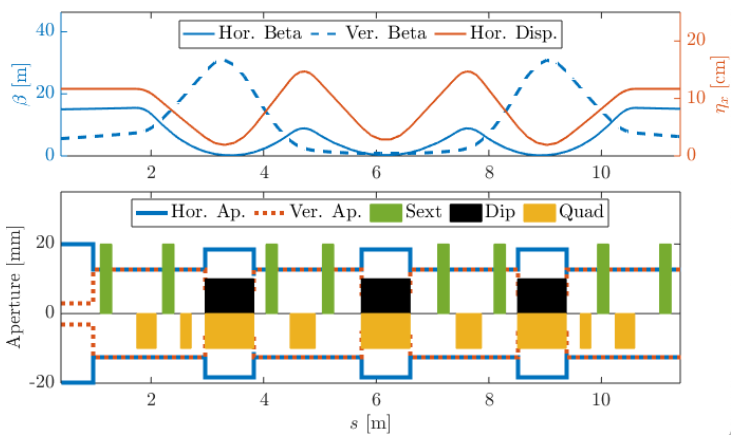}
	\caption{\label{fig:Lattice}
	Lattice and magnet distribution of an arc in the ALS-U AR.
	The ring is composed of $12$ such arcs.
Shown are the beta and the dispersion functions (top), the aperture model and
the distribution of magnets (bottom).}
\end{figure}

The ALS-U AR is essentially a slightly smaller version of the existing ALS twelve period triple-bend acromat.  Its basic parameters, along with those of the existing ALS, are shown in Table \ref{tab:AR}.  The optics and layout through one 
arc are shown in Fig.~\ref{fig:Lattice}.  To adapt the ALS layout to the AR layout, the length of the straight sections
was shrunk from $9.386$ m to $8.762$ m and the arcs were shrunk from
$7.014$ m to $6.415$ m by reducing the magnet spacing.  Nominal bend, quadrupole, and sextupole magnet
lengths are the same between the ALS and AR.  These adjustments bring the total circumference 
down from $196.805$ m to $182.122$ m which makes the AR fit neatly along the inner tunnel wall.  
The AR utilizes a non-swept gradient dipole design, as does the ALS.
The beam energy is increased from $1.9$ GeV
to $2.0$ GeV to match that of the ALS-U storage ring.  
The AR emittance of $1.8$ nm is comparable to that of the ALS. 

To preserve egress, the AR will be mounted close to the tunnel ceiling, about $2$ m above 
the floor and about $0.6$ m above the plane of the storage ring.  The AR magnet stands are a combination 
of floor and wall attachments.  The elevated installation and limited space inside the tunnel
places a premium on reducing the size and weight of the magnets, and so the 
vacuum chamber dimensions were shrunk, bringing the pole tips closer to beam, achieving the same 
gradients with less bulky magnets.  The arc and straight chambers are round and $14.2$ mm in radius.  The dipole
chambers are elliptical with a $20$ mm axis in the horizontal and $7.28$ mm axis in the vertical.

The AR-to-SR injection swap is expected to occur approximately every $30$ seconds.  A long beam lifetime in
the AR is not necessary and so dynamic aperture and injection efficiency are prioritized when optimizing the 
AR.  The AR dynamic aperture thus
extends to the physical apertures.  Simulations taking imperfections and correction algorithms into account 
bring the expected usable dynamic aperture close to linear acceptance.

\begin{figure}[b]
\centering
	\includegraphics[width=0.45\textwidth]{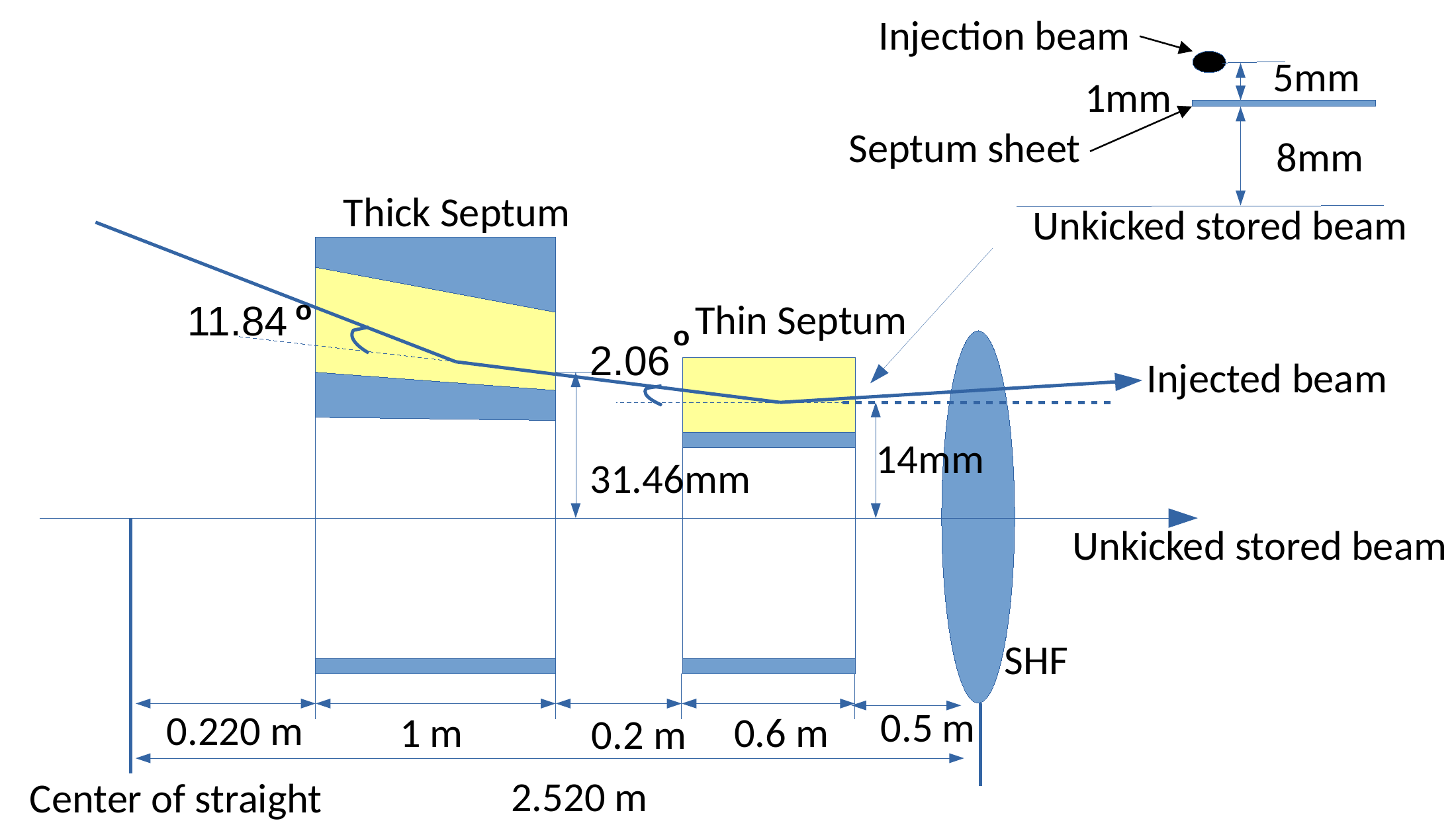}
 \caption{
 \label{fig:septum}
 The layout diagram of the BTA thick and thin septa in sector 1. The separation between the injected 
 beam and unkicked stored beam is $14$ mm and the thickness of the thin septum blade is $1$ mm.}
\end{figure}

Transfer lines intersect the AR along $3$ consecutive straights.  Sector $12$ is the take-off of the accumulator-to-storage ring (ATS) transfer line; the straight section has to accommodate the dipole kicker and pulsed thin septum required for extraction from the AR into the ATS in addition to the first pre-kicker for AR injection from the booster.   Sector $1$ houses the pair of booster-to-accumulator (BTA) injection septa, which are depicted in Fig.~\ref{fig:septum}; immediately upstream in the same straight section is the second pre-kicker for
AR injection from the booster.  The sector $2$ straight contains the landing point of the storage 
ring-to-accumulator (STA) transfer line.  It has to accommodate the STA pulsed thin septum, as well as both the 
dipole kicker for injection from the SR and the main kicker for AR injection from the booster.
The close proximity of the three transfer lines and 
injection/extraction components posed a design challenge, requiring careful consideration of the layout and close interaction between the beam physics and engineering groups.

% moved here from Section III (3DK injection scheme)
The arrival of the injected bunches from the booster is timed to replenish the bunches of 
the depleted train 
in a top-off manner.  In normal ALS-U operations we expect that a train 
circulating in the SR will lose about $10\%$ of its charge before being swapped out.  The beam coming out of the 
booster is $4$ pulses long, with a separation of $3$ empty buckets between each pulse.  The total 
charge in the $4$ pulses is approximately $0.5$ nC, or about $10$\% of the charge that 
needs to be replenished, implying that about 10 injection shots will be needed to replenish 
the circulating train before swapping it back to the SR. 

On-axis injection from the SR into the AR is more straightforward than injection from the booster; this and
extraction from the AR into the SR are not in the scope of this paper.

%\begin{figure}[h]
%	\centering
%	\includegraphics[width=0.48\textwidth]{bad-Sec2-placeholder.png}
%	\caption{\label{fig:sec2}
%	\red{placeholder} Sector $2$ layout containing the STA landing point and injection dipole kicker.
%	Pulses injected in Sector $1$ are kicked to stable trajectories by the injection dipole kicker.}
%\end{figure}

%\begin{table}[htb]  \centering
%   \caption{\label{tab:aperture} \mv{[We may want to reconsider whether to keep this Table;  detailed information on vacuum chamber aperture doesn't seem needed  once the point is made in the text that by design the vacuum chamber is quite narrow ] }  AR vacuum chamber design (inner values) and the corresponding values for the aperture model. The given values indicate the radius $r$ for circular (circ.) apertures, the half-axis radii $r_x/r_y$ for elliptical (el.) apertures, and the ranges $[-x,+x]/[-y,+y]$ for rectangular apertures (all values in \SI{}{\mm}). }
%\begin{ruledtabular}
%\begin{tabular}{l|c|c|c}
%					            &profile & actual dims.		& as modeled	\\ 
%\hline
%Dipole			        & el.    & 20 / 7.28				& 18.3 / 5.6		\\
%Straights		    	& circ.  & 14.2				    & 12.6			\\
%Arcs  			        & circ.  & 14.2					& 12.6			\\
%Dip. Rad. Shield 	    & el.    & 12.3/14.2				& 10.9/12.6		\\
%Septum to 1st Dipole  	& el.    & \red{22 / 7.28}		& 20.7 / 5.1		\\
%1st to 2nd Dipole 	    & el.    & \red{16.2 / 14.2}		& 14.6 / 12.6		\\
%Septum 			        & rect.  & [-25,8]/[-25,25]		& [-25,8]/[-25,25]	\\
%Fast kicker 		    & rect.  & [-20,20]/[-3,3]		& [-20,20]/[-3,3]		\\
%   \end{tabular}
%   \end{ruledtabular}
%\end{table}

\section{3DK Injection Scheme}
\label{sec:3DK}

\begin{figure*}[ht]
\centering
	\includegraphics*[width=1.0\linewidth]{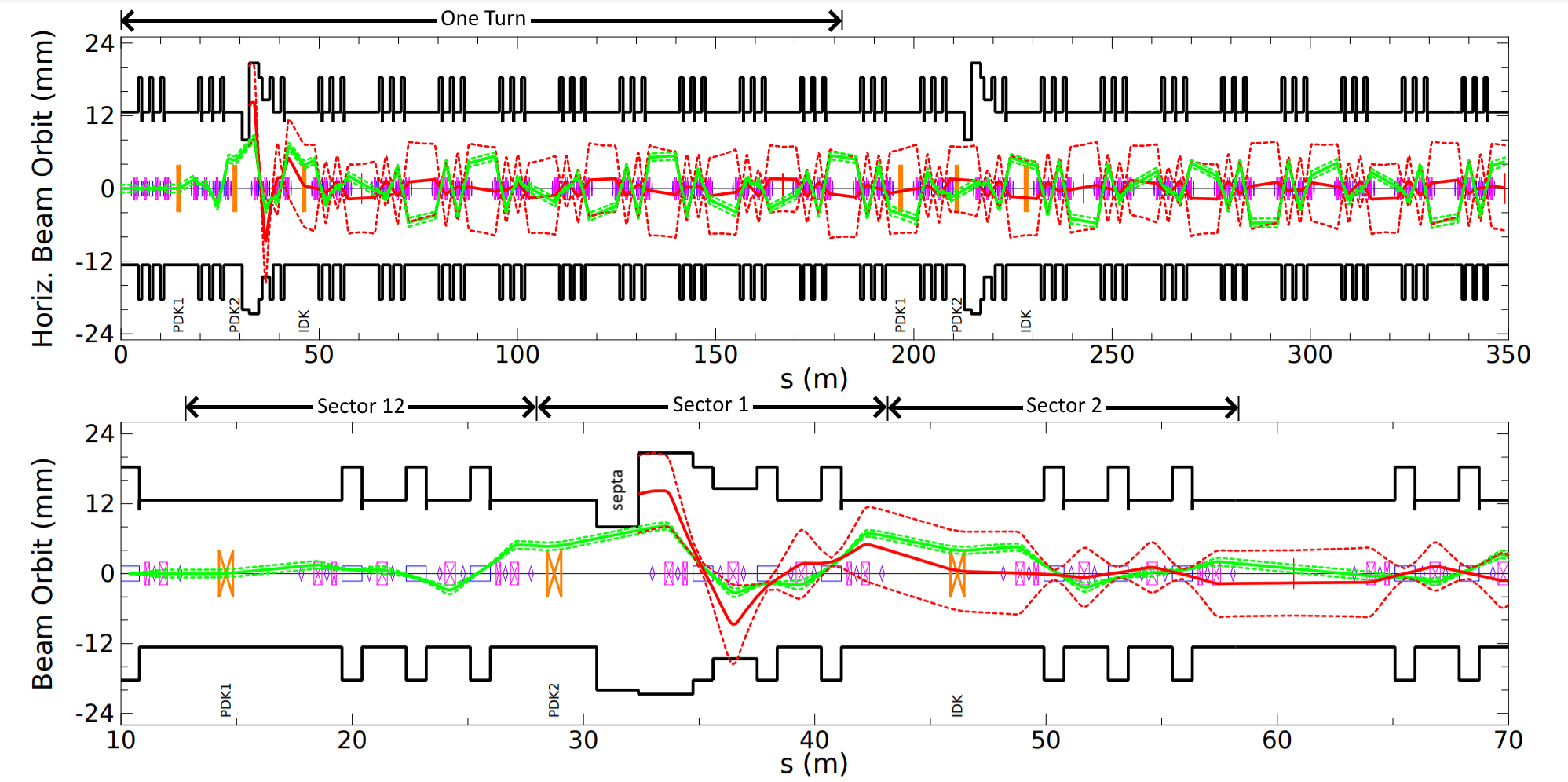}
 \caption{\label{fig:3DK-balanced}
 Trajectories of the stored (green) and injected (red) beams during an injection cycle for the 
 ``balanced'' mode kicker strength settings, also referred to as ``Mode A''.  Solid black profile
 indicates horizontal physical aperture.  Dashed lines 
 delineate $3\sigma$ of the beam size.  Kickers are labeled and represented by orange bow-ties.  The first pre-kicker
 is located at the start of sector $12$.  The second pre-kicker and injection septum are 
 located in sector $1$.  The main injection kicker in sector $2$ kicks both the stored and injected 
 particles.}
\end{figure*}

The AR layout along with injection elements and beam trajectories during an injection cycle
are depicted in Fig.~\ref{fig:3DK-balanced}.
At the start of a BTA injection cycle, the stored bunch train is given an outward ``pre-kick'' 
at the upstream end of the sector $12$ straight.  A second outward pre-kick is applied at the
upstream end of the next straight, which is sector $1$.  The injection septum is at the downstream end 
of sector $1$.  Injected particles make a large excursion through the sector $1$ arc.  
The main injection kicker, located at the downstream end of sector $2$,
applies an outward kick to both the stored and injected particles, leaving both on 
trajectories that damp with very little particle loss.  The strengths of the $3$ kickers and injection
septum are adjusted to control the injection transients and are shown in Table \ref{tab:inj-modes}.
The kickers' flat tops span the entire bunch train and are timed to the train head, 
irrespective of which buckets are being injected into.

\begin{table*}[]
\caption{Injection modes achieved by adjusting kickers and septum strength.  Mode A achieves
a high injection efficiency with a small stored beam transient.  
Mode B reduces the stored beam transient to mitigate short range wake fields, at 
the expense of greater losses on the BTA-side septum sheet and due to injected particles exceeding 
the acceptance.  Mode C recovers high injection efficiency by shrinking (mismatching) the 
horizontal $\beta$ of the injected beam.}
\label{tab:inj-modes}
\begin{tabular}{c|l|c|c|c|c|c}
\hline
\hline
                    &&\thead{On-axis\\injection}&\thead{Closed\\stored-beam\\bump}&\thead{Mode A}&
                    \thead{Mode B}&
                    \thead{Mode C}\\ 
\hline
\multirow{4}{*}{\thead{Injection element\\strengths}}
&Thin septum (mrad)           &$35.82$&$35.85$&$35.85$&$35.72$&$35.67$\\
&Pre-kicker 1 (mrad)     &$0.365$&$0.489$&$0.383$&$0.417$&$0.417$\\ 
&Pre-kicker 2 (mrad)     &$1.09$&$0.759$&$1.04$&$0.933$&$0.933$\\
&Inj. kicker (mrad)      &$1.21$&$0.613$&$1.11$&$0.987$&$0.987$\\
\hline
\multirow{4}{*}{\thead{Post-injection\\cycle residuals}}
&Stored beam rms (mm) &$2.88$&$0.0$&$2.44$&$1.86$&$1.86$\\
&Stored beam max (mm) &$6.32$&$0.0$&$5.39$&$4.18$&$4.18$\\
&Injected beam rms (mm) &$0.0$&$3.74$&$0.734$&$1.02$&$0.996$\\
&Injected beam max (mm) &$0.0$&$8.16$&$1.59$&$2.21$&$2.17$\\
\hline
\multirow{2.5}{*}{\thead{Optics at\\septum exit}}
&$\beta_x$ (m) &$14.82$&$14.82$&$14.82$&$14.82$&$11.33$\\
&$\alpha_x$ &$-0.139$&$-0.139$&$-0.139$&$-0.139$&$-0.275$\\
\hline
&BTA-side septum losses (\%)     &$0.89$&$0.89$&$0.89$&$1.05$&$0.48$\\
&Capture losses (\%)         &&&&&\\
&~~~~no wakes            &$-$&$66.3$&$0.36$&$0.91$&$0.46$\\
&~~~~with wakes          &$-$&$-$&$16.4$&$6.8$&$6.9$\\
&~~~~with wake mitigations &$-$&$-$&$0.7$&$1.1$&$1.0$\\
&Stored Beam Survival (\%) &$99.9$&$99.9$&$99.9$&$99.9$&$99.9$\\
\hline
\hline
\end{tabular}
\end{table*}

The vacuum chamber between the injection point and the second downstream dipole is horizontally widened to accommodate the large excursion of the injected-beam trajectory but no special-aperture magnets are required.

The injected bunches arriving from the pulsed thin septum are offset $14$ mm outward from the nominal 
chamber center, as depicted in Fig.~\ref{fig:septum}.  
The transverse position of the septum 
was arrived at by considering both the limitation the septum places on 
dynamic aperture and the constraint it imposes on the beam-trajectory oscillation amplitude 
during the post-injection transient.  The distance 
of the injected beam centroid to the septum sheet was arrived at by considering particle loss on the septum sheet 
and the amplitude of the injected particles.

The starting point for determining the optimal septum angle is found by adjusting the angle of the 
injected particles to create a zero-crossing at the main injection kicker.  The septum angle is 
refined along with the strengths of the $3$ kickers by optimizing amplitudes of the stored and injected 
transients in action/angle coordinates.
The optimum is that which minimizes the action while preserving minimum distances from the 
limiting physical apertures and tolerating no losses out to $3\sigma$ of stored beam.

%The action of the beam centroid is defined as
%\begin{equation}
%J^2_x = \gamma_x^2\left<x\right>^2+2\alpha_x\left<x\right>\left<x'\right>+
%\beta_x\left<x'\right>^2\textrm{.}
%\end{equation}
%\mv{[Do we need to  report this equation? In any case it should be $J_x$ not $J_x^2$. ] }

The optimum does not result in a zero-crossing
at the injection kicker.  In fact, placing either the injected beam or the stored beam
onto the reference orbit at the end of the injection kicker results in unacceptably large trajectories
of the off-axis beam.
Instead, the optimum involves a trade-off between reducing the transient amplitudes of the injected particles 
and the stored beam.  

These options are explored in Table \ref{tab:inj-modes}, which presents the thin septum and kicker strengths for different injection modes, as well as the maximum amplitude and rms of the horizontal injection transient for the stored and injected beams. The ``on-axis injection mode,'' as was done at MAX IV \cite{LEEMANN2012117}, is expected to be
useful during early commissioning.  It will place the injected beam on-axis, but the stored beam, 
if present, will be left with a large oscillation amplitude that limits imperfection tolerance and generates
large short-range wake fields.  
The ``closed stored-beam orbit bump mode'' eliminates the stored beam injection transient, but
achieves only $63$\% injection efficiency.  
It is included here to illuminate the bounds of the parameter space.
The remaining three modes indicate different strategies to optimize performance for
small but non-zero residual amplitudes of the stored and injected beams.
Mode A, or ``balanced mode,'' represents a compromise solution between the two previous
modes and maintains near-$100$\% injection
efficiency while reducing the amplitude of the stored beam oscillations, with an injected beam
that is matched to the storage ring.
For this mode, the transient of the stored beam is allowed to be larger than that 
of the injected beam as it has a much smaller beam size than the injected particles.
The trajectories in Fig.~\ref{fig:3DK-balanced} are based on Mode A.

The parameters of the injection elements are further refined after considering short-range 
transverse wake fields.  
Because of the non-zero injection transient, the stored beam generates relatively strong
wakes which can sweep some of the injected particles into the vacuum chamber.  This is described in
more detail in Sec.~\ref{sec:short-range-wakes}.  When taking short range wakes into account, 
it is found that a smaller perturbation to the stored beam improves the overall injection 
efficiency, even though the injected bunch has a larger initial offset and suffers slightly larger
losses initially.

Mode B brings the injected beam closer to the septum as it enters from the BTA thus requiring a weaker
main injection kick and allowing for a smaller stored beam transient at the expense of slightly reduced injection
efficiency.  The injected beam losses to the machine acceptance increase from $0.36$\% to
$0.91$\%, while the losses due to wake fields decrease from $16.4$\% to $6.8$\%.
BTA-side septum sheet losses also increase slightly from $0.89$\% to $1.05$\%.

Mode C has the same kicker settings as Mode B but decreases $\beta_x$ to better fit the 
injected beam into the AR acceptance available after taking the septum sheet into account,
similar to the technique applied in \cite{streun-inj-mismatch}.  
This differs from Modes A and B where the beam distributions are matched to the ring.  
Reducing $\beta_x$ shrinks the injected beam horizontally, reducing losses along the septum sheet.  
After adjusting the
septum strength to re-center the injected beam within the available machine acceptance, the centroid
is brought closer to the septum, thereby reducing the injection transient.  BTA-side septum
sheet losses are reduced to $0.48$\%, and losses to the acceptance are reduced to 
$0.46$\%.

\begin{figure}[b]
\centering
\includegraphics[width=0.42\textwidth]{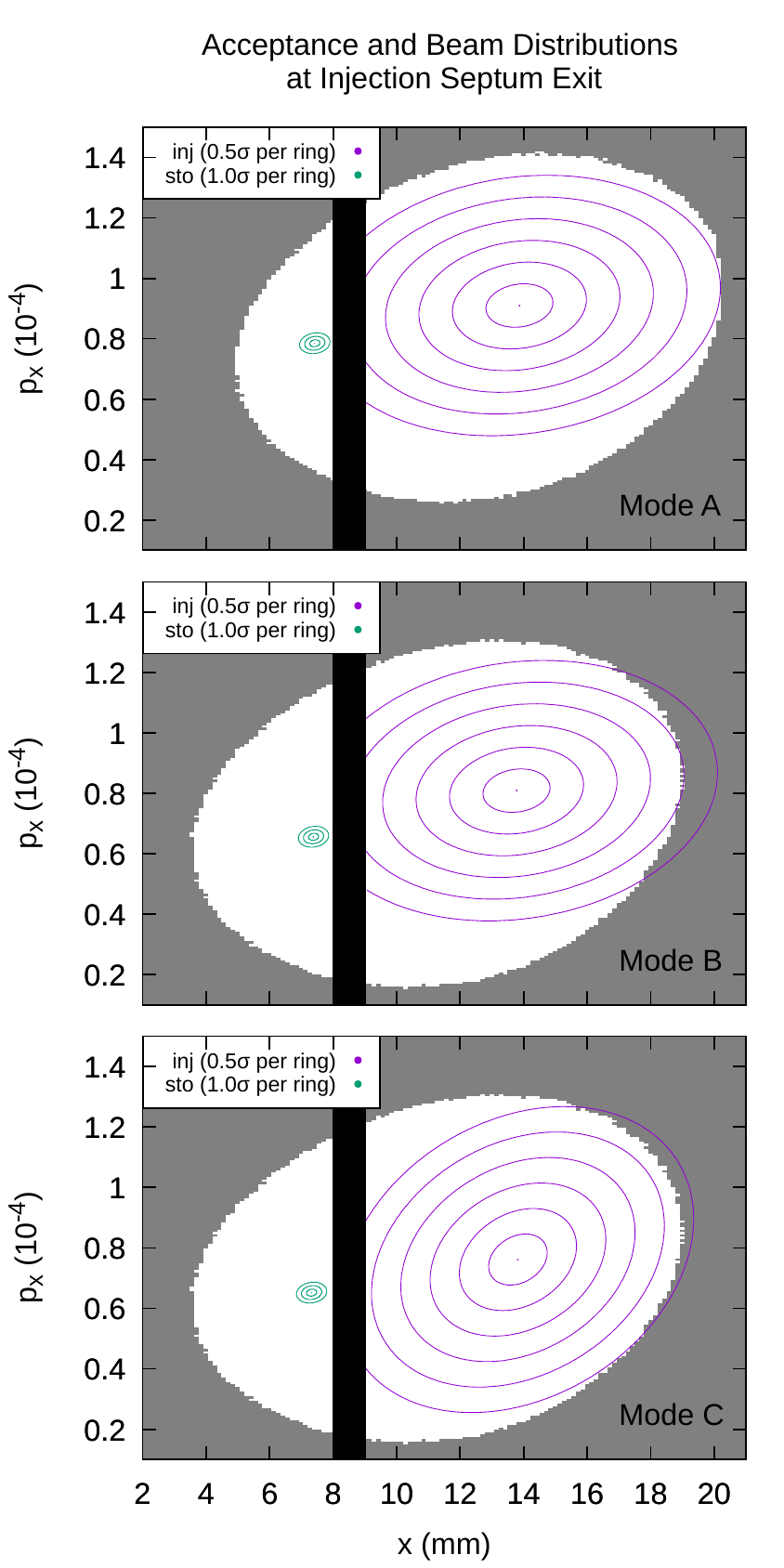}
\caption{Injected and stored beam profiles out to $3\sigma$ at septum exit during an injection 
cycle for the three injection modes.    
The white region is the acceptance in $x$-$x'$ space at the septum exit.
The septum is drawn in black and is $1$ mm thick with its inward edge $8$ mm from the stored beam reference trajectory.
The septum image was obtained by tracking.
The parameters of the three modes are detailed in Table \ref{tab:inj-modes}.
 \label{fig:da-dists}}
\end{figure}

Fig.~\ref{fig:da-dists}
shows the DA of the latter three modes viewed from the septum exit during an injection cycle, 
along with the injected and stored particle distributions, and also the thin septum.  Both the inward
and outward profiles of the thin septum were determined by particle tracking.  Comparing
Mode B to Mode A, the acceptance appears to move since the acceptance itself
relative to the septum is affected by the choice of main injection kick.
%After taking wake fields and an ensemble of error realizations and corrections into account, 
%modes B and C consistently yield greater than $95$\% injection efficiency. 
%Note that Mode A can achieve similar performance after applying wake mitigations, namely,
%increasing the AR chromaticity or delaying the arrival of the injected particles by roughly $10$\% of the
%RF bucket.

%\begin{figure*}
%    \centering
%    \includegraphics[width=\linewidth]{bad-painting.png}
%    \caption{\label{fig:beam-painting}
%    \red{placeholder} The ``beam painting'' filling pattern necessitated by the bunch structure    coming out of the  booster.  A booster shot consists of $4$ bunches separated by $8$ ns.  The     total current of the shot is $0.5$ mA, distributed amongst the $4$ bunches approximately     as $7$\%, $33$\%, $40$\%, and $20$\%.} \end{figure*}

%AR designcurrent.  A ``beam painting'' schedule, depicted in Fig.~\ref{fig:beam-painting} is 
%utilized scheme  to evenly fill the AR train.}

In producing the nominal operational mode and its two variants, we conclude that the 3DK scheme 
appears to have enough flexibility to accommodate perturbations and offers various 
optimization opportunities.

\section{BTA layout design and optics matching}\label{sec:BTA}

Transport from the booster to the AR is through the
Booster-to-Accumulator (BTA) transfer line. The BTA utilizes about 2/3  of the existing booster-to-ALS storage ring (BTS) transfer line. For early commissioning a slow  bending switch will be installed for on-demand steering of the beam into the new BTA branch while ALS can continue normal operation.   
%To shorten the “dark time” and minimize impacts on beamline users during the ALS upgrade, the 
%accumulator ring will be installed and
%commissioned while the ALS is still in operation. This
%requires a coexistence of injections to both the accumulator
%ring and the ALS storage ring. To facilitate this, a “switcher bend” will be installed in the 
%existing Booster-to-Storage ring (BTS) transfer line. This switcher bend will branch out the 
%BTS line to the new BTA line, allowing for switching the injections between the ALS for 
%normal user operation, and the accumulator for commissioning.
% As menti, the accumulator is mounted on the inner shielding wall of the storage ring tunnel at a different elevation level from the booster ring, as shown in Fig.~\ref{fig:machine}. 
%

\begin{figure}[!t]
\centering
	\includegraphics[width=0.48\textwidth]{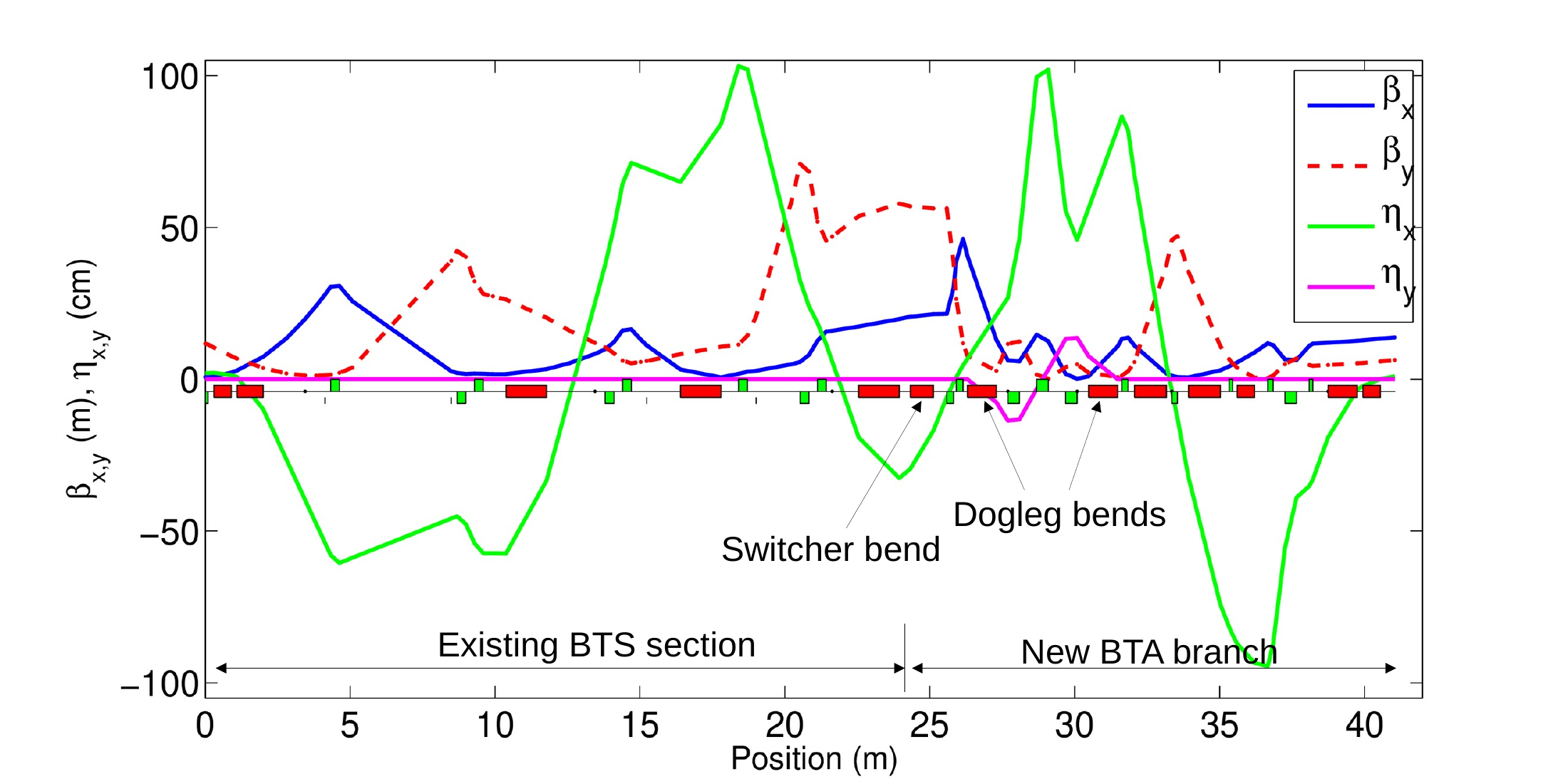}\\
	\includegraphics[width=0.45\textwidth]{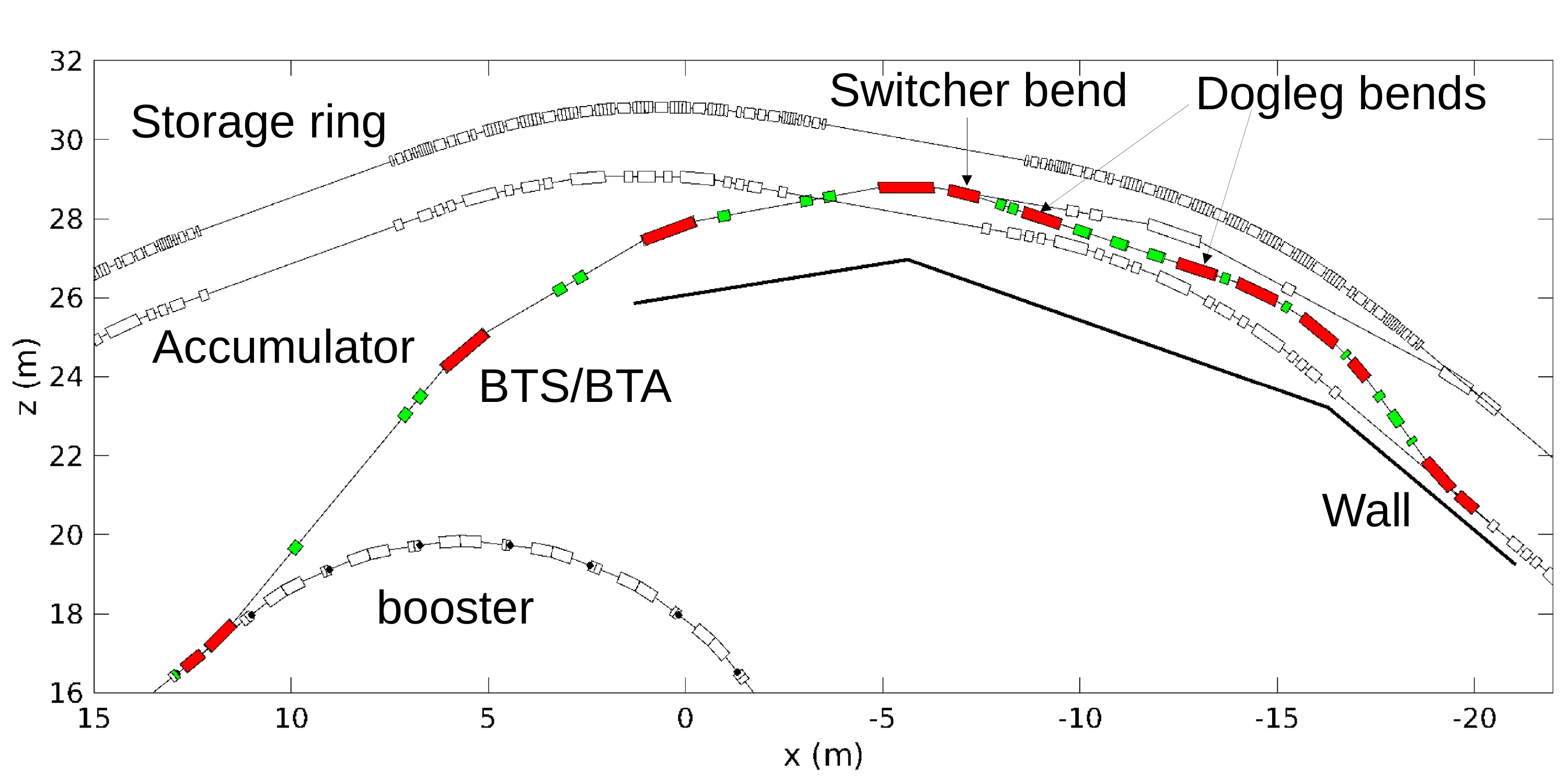}
 \caption{\label{fig:BTA_twiss_env}
 Twiss functions (a) and layout(c) of the BTA transfer line.}
\end{figure}

To accommodate the different elevation between booster and AR (see Fig.~\ref{fig:machine}), the BTA includes  a vertical two-bend achromatic  dogleg  \cite{Sun:IPAC2019-WEPGW111}, where  the achromatic condition is met with two quadrupoles between the two dogleg bends to generate the required  vertical $\pi$  phase advance. The BTA terminates into a pair of pulsed thin and thick septa providing a combined 12~deg (horizontal) bending. 
%These septa are located at the far end of the injection straight (straight 1) as shown in Fig.~\ref{fig:septum} %to reduce the bending angle requirement for these septa.
Since the space is limited, thick and thin septa are located at the far end of the injection straight (straight 1) as shown in Fig.~\ref{fig:septum} to increase the distance from the last bend of the BTA to the septa and lower the incoming angle into the septa, therefore reducing the required kick angles from septa.

The main design difficulty posed by the BTA  is the  avoidance of  interference between the BTA and other systems' components in the tight and congested injection region while meeting the layout and lattice functions' matching boundary conditions. 
%
%To overcome these challenges, the BTA layout is optimized in the ALS global coordinate system by adjusting the bending angles of dipoles and the lengths of drift spaces along the line.  Once the layout is determined, 
%
Linear-optics  matching  was performed with constraints on beam sizes along the transfer line to minimize vacuum chamber and magnets' aperture.  The decoupling of horizontal and vertical planes provided by the dogleg simplifies the matching problem but we found it useful to deploy a Multi-Objective Genetic Algorithm (MOGA)~\cite{996017}  to optimize the design. 
%
%Since the vertical dispersion bump is localized to the dogleg, the vertical achromatic condition for the dogleg can be matched independently from the horizontal optics. The optics matching is carried out using a Multi-Objective Genetic Algorithm (MOGA)~\cite{MOGA}. All the quadrupoles in both the existing BTS line and the new BTA branch are used as matching knobs.  The optimization objective is to match the Twiss functions of the BTA transfer line to the accumulator ring at the injection point. The beta and dispersion functions are constrained to minimize the beam size throughout the transfer line. 
%
The optics functions and beam-size envelopes of the baseline lattice are shown in Fig.~\ref{fig:BTA_twiss_env}. 
The desired Twiss functions at the end of the transfer line are obtained and the beam sizes are 
below $4$ mm throughout, which satisfies the design requirements.

\section{Injection efficiency analysis and robustness against lattice errors}\label{sec:imperfections}

The injection efficiency is evaluated using AR and BTA lattices obtained after undergoing a process of simulated commissioning~\cite{PhysRevAccelBeams.22.100702}.

Optimizing the injection efficiency in the presence of errors requires small adjustments to the nominal kick angles, with the exact value depending on the error realization. In simulations and eventually in operation, the  optimization can be accomplished by running parameter scans in a neighborhood of the kick-angle nominal settings. 

While one would have the freedom to adjust all three DKs independently, in the simulations we used a simplified setup in which both pre-DKs are varied by the same amount, thus resulting in a 2D optimization problem. We found that this setup gave satisfying results while significantly reducing the computation time compared to a full 3-parameter scan.

\begin{figure}[hb]
\centering
	\includegraphics[width=0.48\textwidth]{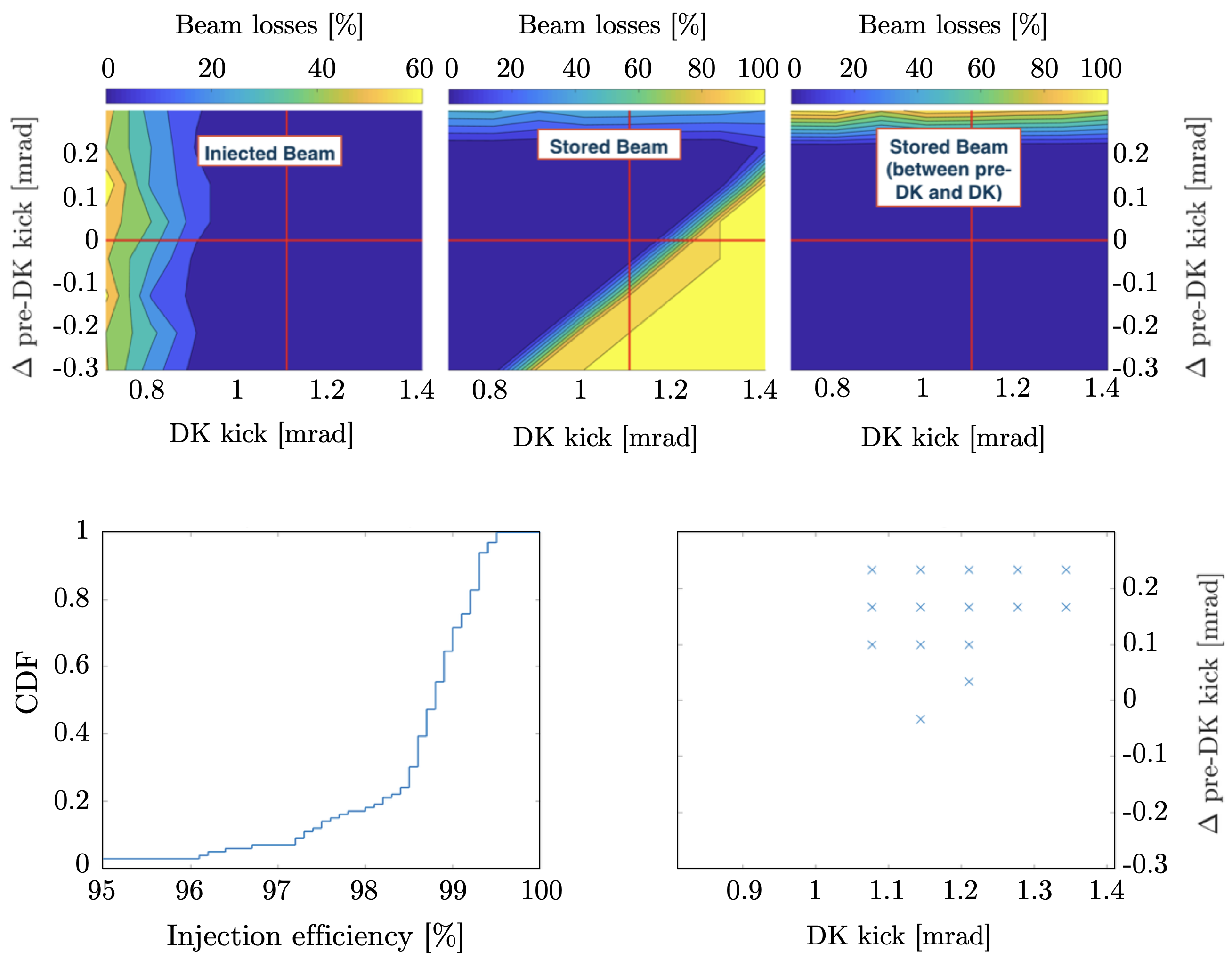}
 \caption{\label{fig:kickscan}
 Top: particle losses in the injected (left) and stored (center and right) beam  vs. kickers' angles for the ideal lattice. The pre-kickers' kick (vertical axis)  is reported relative to Mode-A settings (red lines).  
 Bottom: for each one  of 100 lattice-error realizations, a DK scan is carried out on a grid to identify the setting yielding the highest injection efficiency. The ensemble of the  best DK settings is reported in the right figure. The left figure is the injection efficiency CDF. }
\end{figure}

In a preliminary study we took a first cut at the parameter space by delimiting the region that  exhibits no or minimal losses in both injected and stored beam in the absence of errors: see  the top images in Fig.~\ref{fig:kickscan}. 
In the simulations the injected and stored bunches are represented with $1000$ particles and tracked for $1000$ turns.  

\begin{table}
\centering  
   \caption{\label{tab:error_sources} Imperfections assigned to injection elements in the commissioning simulations. The error distributions are Gaussian and truncated at $2\sigma$.  The stability errors are relative to the set point, while the roll error is absolute.}
   %\begin{ruledtabular}
\begin{tabular}{l|c}
\hline
\hline
Error Type                          & Distribution Width	\\
\hline
Septum stability             & $2.5\cdot10^{-4}$	\\ 
Dipole kicker stability             & $1\cdot10^{-3}$	\\ 
Dipole kicker roll                 & \SI{4}{\mrad}	\\ 
\hline
\hline
   \end{tabular}
   %\end{ruledtabular}
\end{table}

\begin{figure}[h]
\centering
	\includegraphics[width=0.48\textwidth]{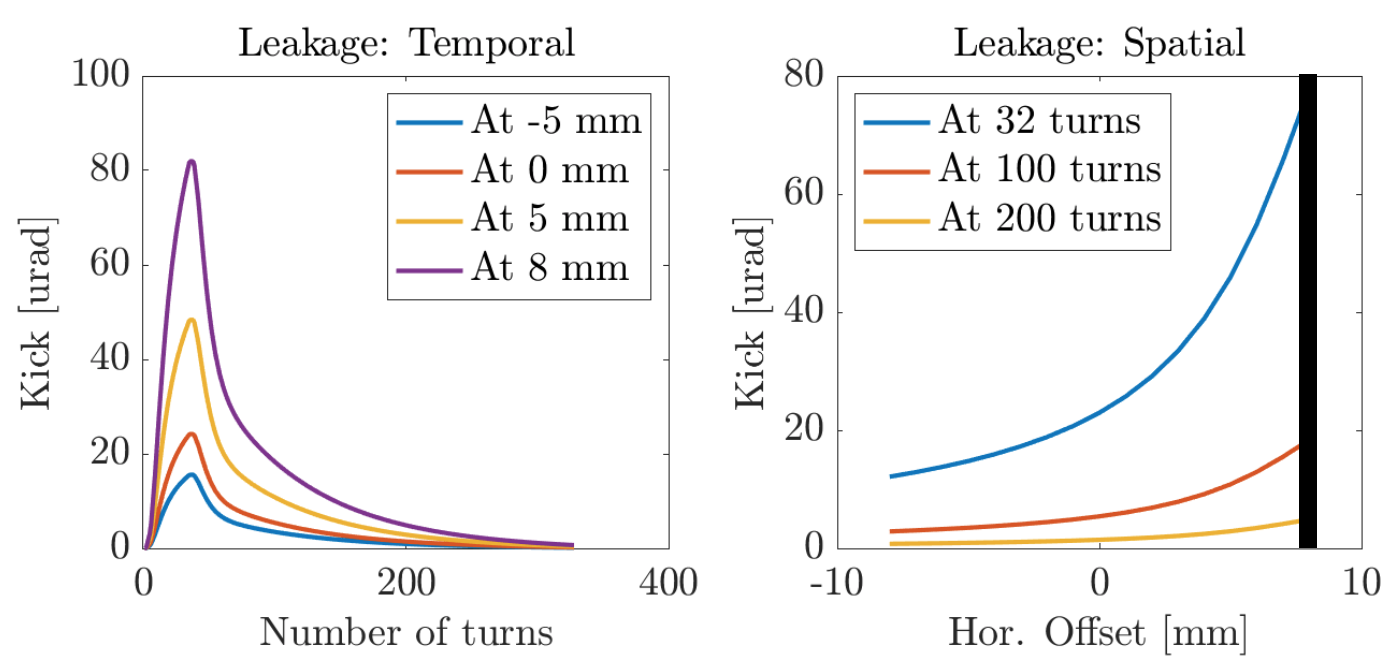}
 \caption{\label{fig:leakageField}  Temporal  and spatial profiles of the thin-septum leakage field as included  in the injection efficiency studies. The left plot shows the integrated kick  at various  horizontal coordinates as a function of turns. Conversely, the right plot shows the integrated kick at various times as a function of a particle horizontal coordinate. }
\end{figure}
\begin{figure}[!t]
\centering
	\includegraphics[width=0.48\textwidth]{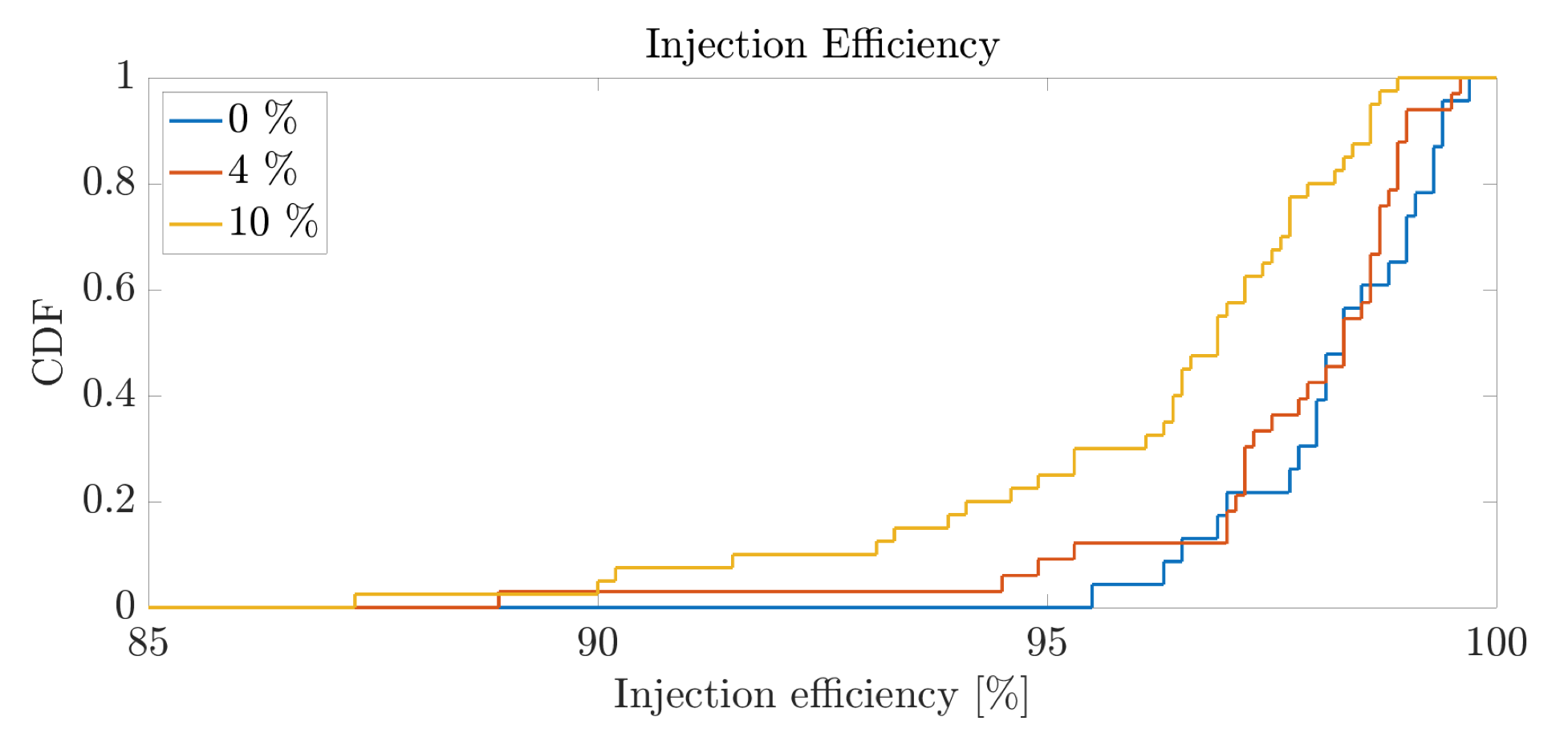}
 \caption{\label{fig:injectEffic}
Sensitivity of the injection efficiency to the non-uniformity of the injection kickers’ pulse 
profile. The data sets correspond to three amounts of pulse profile slope indicated in the plot key. 
The simulation is conducted with a complete set of errors including shot-to-shot variation of the kick and septa strengths, as well as septum leakage fields.}
\end{figure}

As the best combination of kicker settings may vary slightly depending on the specific random error realization and correction, a 2D kicker scan was carried out for each of the $100$ random lattice-error realizations considered. For each lattice-error realization, the kickers' settings that induce no stored-beam particle losses were selected and among these 
were identified the one kickers' setting that gives the highest injection efficiency. The set of those best 100 settings  (one per lattice)  is shown in the bottom-right image of Fig.~\ref{fig:kickscan}  (only $15$ distinct data points appear, as a  single kickers’ setting will in general represent the optimum for several lattice-error realizations.)  The bottom-left picture in Fig.~\ref{fig:injectEffic} is the CDF of this set, showing that about half of the  lattices have injection efficiency larger than $98.5$\%.

We then proceeded by refining  the simulations to include a more complete set of errors and perturbations.  Specifically, to the AR lattice errors we added:   septa and  kickers shot-to-shot variations (Table \ref{tab:error_sources}), non-uniformity of the kicker-pulse temporal profile, and  septum leakage fields.

The non-uniformity of the kicker pulse is simulated by assigning a linear slope to the pulse profile. The slope is defined as the difference between  the  kicks on the first and last bunch of the bunch train relative to the nominal kick. The simulation  of the leakage fields is done by calculating the kick map associated with the field accounting for both the  temporal and spatial field extension, to be applied at every turn before the field has died off,  Fig.~\ref{fig:leakageField}. A result  from these more complete simulations is shown in  Fig.~\ref{fig:injectEffic} demonstrating the sensitivity of injection efficiency to the sloping of the  kickers’ pulse. As before, for each lattice-error realization, the injection efficiency was determined  after performing a 2D scan of the kicker’s amplitude to identify the optimum. Among other things, from  these results we concluded that a $4$\% slope would be acceptable.

%----------- remove this sub-section MV 6/3/21
%\subsection{Constraints on Injection Components}
%The static and pulsed injection components are subject to both time domain and spatial field quality constraints to ensure minimal perturbation to the closed orbit and injection transient. As depicted in Fig.~\ref{fig:septum}, the incoming beam from the booster is first kicked by a  static Lambertson thick septum, then by a pulsed thin septum.  To avoid perturbing the stored beam, the leakage field of the Lambertson is engineered to be less than $X.X$ Gauss, resulting in less than a $X.X$ $\mu$m perturbation to the beam trajectory at the septum.  Figure \ref{fig:leakageField} depicts the spatial and temporal leakage field of the thin septum. The field profile of the pulsed thin septum is such that a reflection is tolerated due to its being confined to  the duration of the long train gap ($26$ bunch train, harmonic number $304$).  Allowing for the reflection simplifies power supply engineering and does not impact the beam.
%\begin{figure}
%    \centering
%    \includegraphics[width=0.48\textwidth]{placeholder_bta_ts.pdf}
%    \caption{Leakage field of the BTA pulsed thin septum versus time.  The circuit is designed
%    such that reflections appear while the $26$-bunch train is many buckets away.}
%    \label{fig:septum-field}
%\end{figure}
%----------- removed  sub-section above;  MV 6/3/21

\section{Single bunch (short-range wake fields)}\label{sec:short-range-wakes}
During injection both the electrons from the booster and the stored bunch onto which they 
are added undergo several millimeter transverse oscillations and this has the potential to induce 
potentially harmful transverse collective effects. The investigation of these effects it the 
topic of this section.

\begin{figure}[!t]
\centering
	\includegraphics[width=0.48\textwidth]{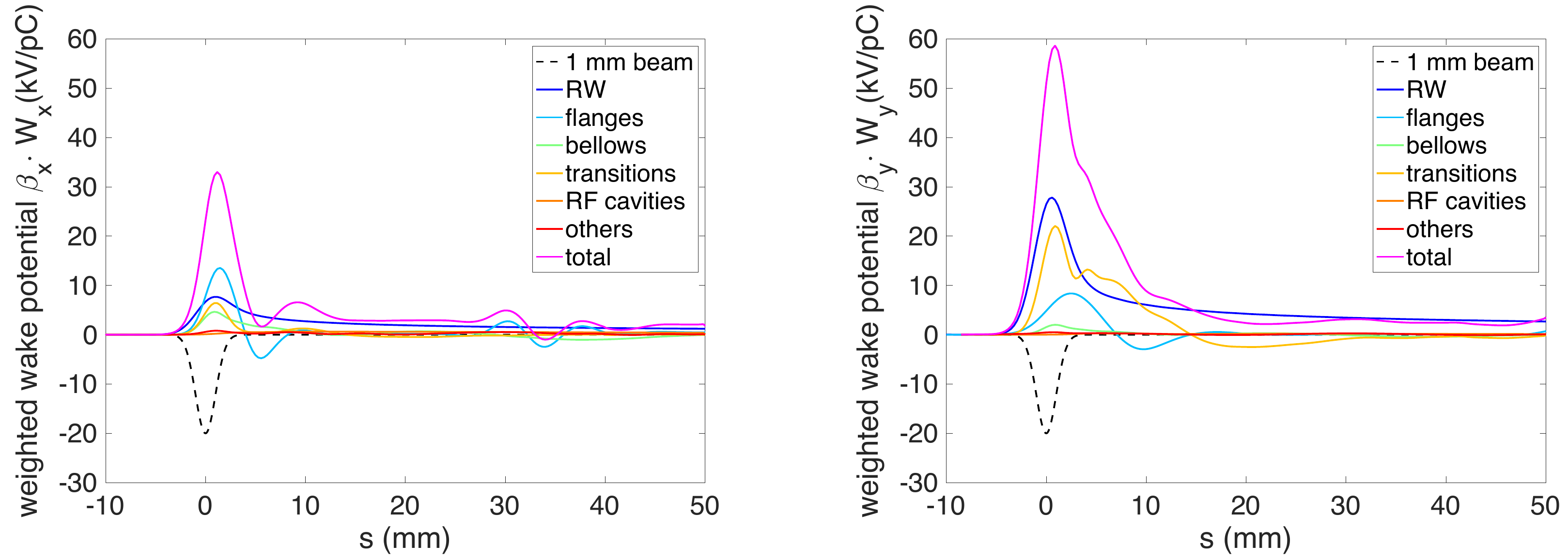}
 \caption{\label{fig:short_wake}
 Transverse short-range wakefields in the AR from individual components and their 
 combined total.  The wake potentials are calculated for a Gaussian beam with rms bunch length 
 of $1$ mm and serve as pseudo-Green functions in beam 
 dynamics simulations. Left: horizontal plane, right: vertical plane}
\end{figure}

The short-range transverse wake fields have been calculated for each vacuum chamber element 
using a detailed model of its geometry.  
Longitudinal wakes are not included in these results, however only the horizontal 
wakes have been observed to have a significant impact on the injection process.
The wakes are applied close to each source point, 
although simulations are in good agreement with simpler calculations that
use beta-function weighting to group together the impact of wake fields
in every sector or even at a single location in the ring.
The threshold for the Transverse Mode-Coupling Instability (TMCI)   is $5.8$~nC, about $8$ times the nominal charge for a single bunch.
Fig.~\ref{fig:short_wake} shows the beta-weighted wakefields from different components
in both horizontal and vertical planes, along with their total.  Wake fields are
calculated for a Gaussian beam with rms bunch length of $1$ mm, and 
serve as pseudo-Green functions in the following beam dynamics simulations.

\begin{figure}[htb]
\centering
	\includegraphics[width=0.48\textwidth]{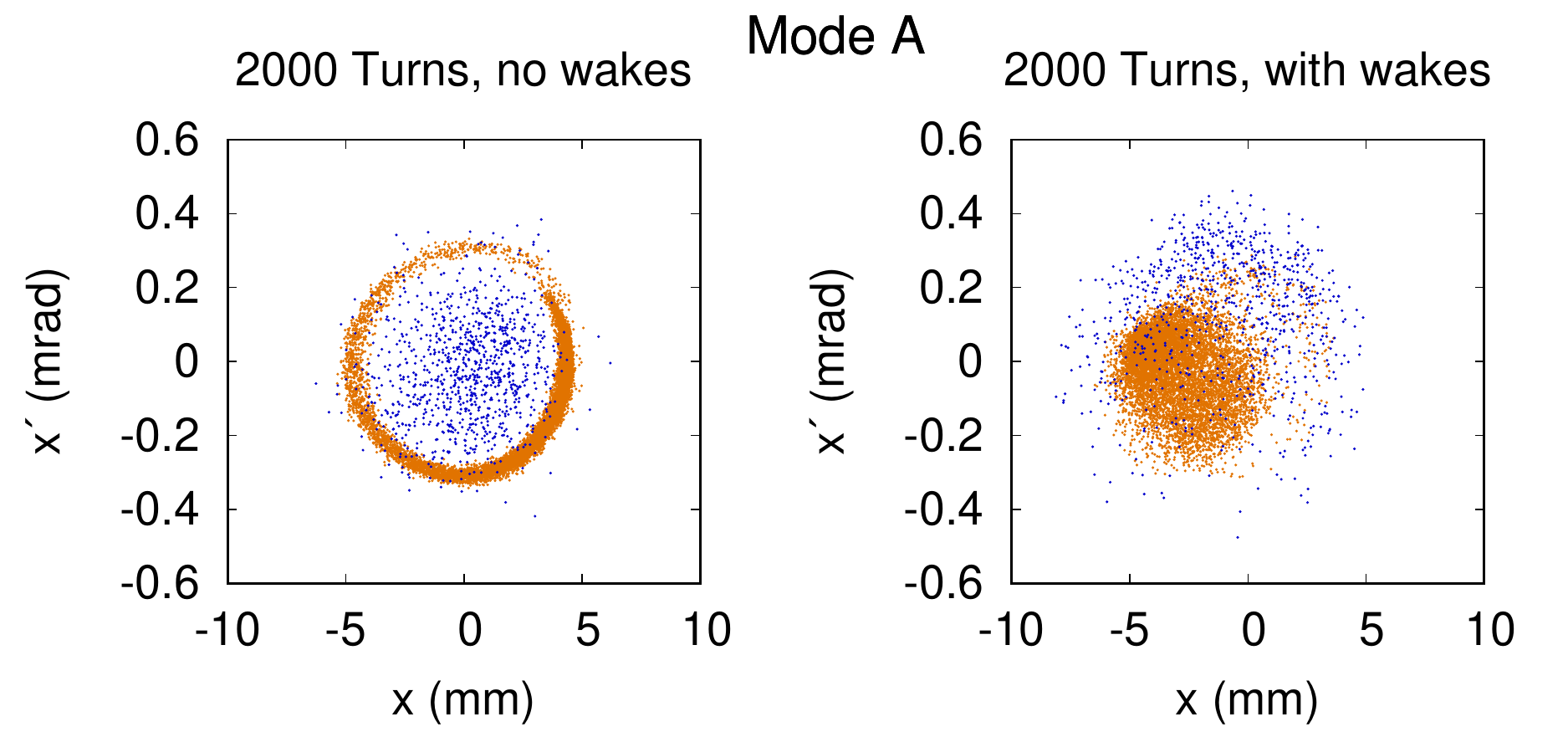}
	\includegraphics[width=0.48\textwidth]{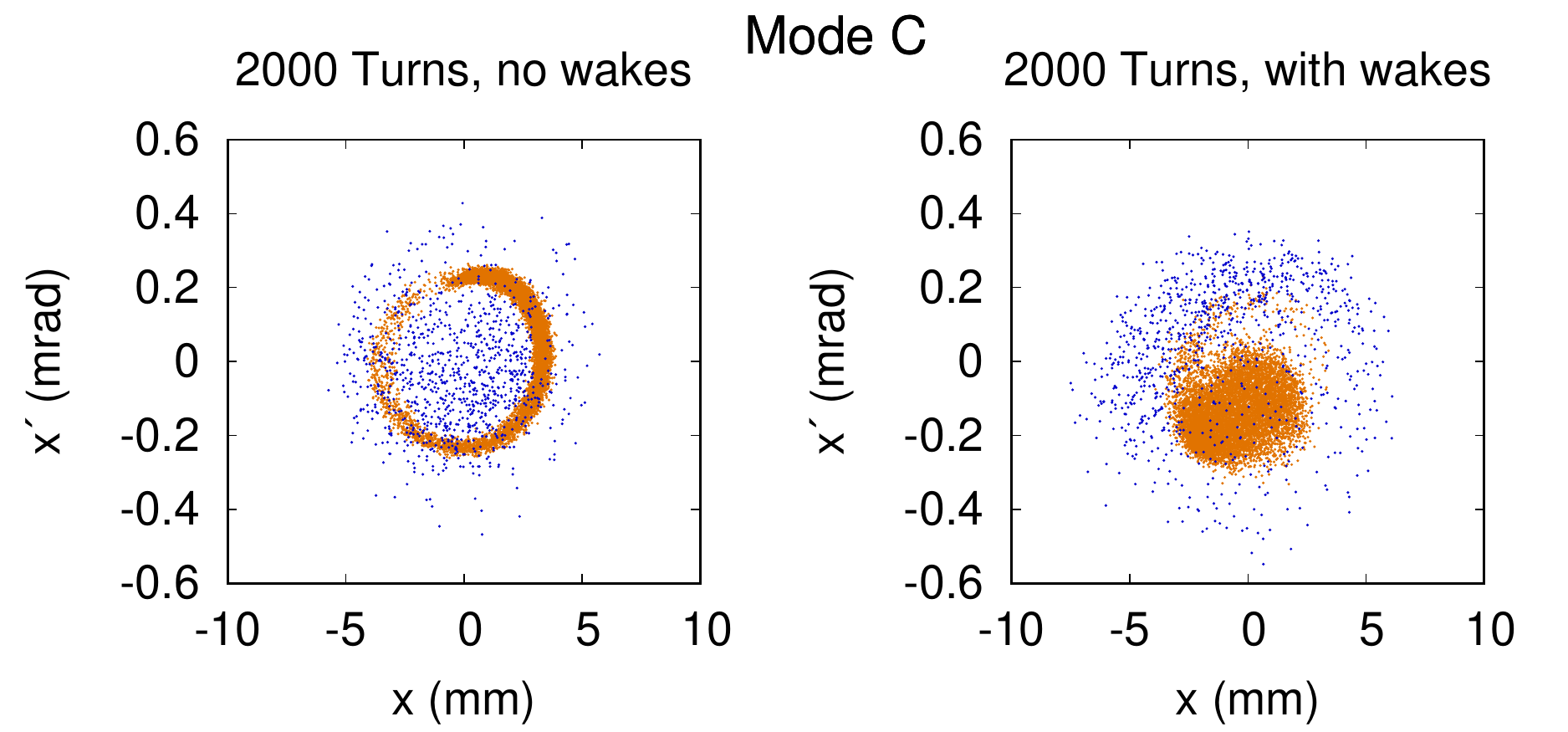}
    \caption{\label{fig:wake_phasespace}
 Horizontal phase space of the stored (red) and injected (green) bunches after $2000$ turns,
 without wakes (left) and with nominal wakes (right).  The upper figures show the results for Mode A, 
 the lower Mode C.}
\end{figure}

The resulting transverse oscillations, especially of the stored bunch (because
it contains about $90$\% of the charge), induce short-range wake fields.
Despite the fact that the total bunch charge is expected to be far from
the threshold for instabilities driven by short-range wakes, the wake fields
do build up in intensity over a time scale of order $1$ ms (roughly $1600$ passes 
around the ring).  Simulations
show that the wake fields are strong enough to confine the stored bunch and
suppress phase decoherence, as shown in Fig.~\ref{fig:wake_phasespace}.
At the same time, the more diffuse injected
electrons increase in both oscillation amplitude and transverse width.

\begin{figure}[ht]
\centering
	\includegraphics[width=0.48\textwidth]{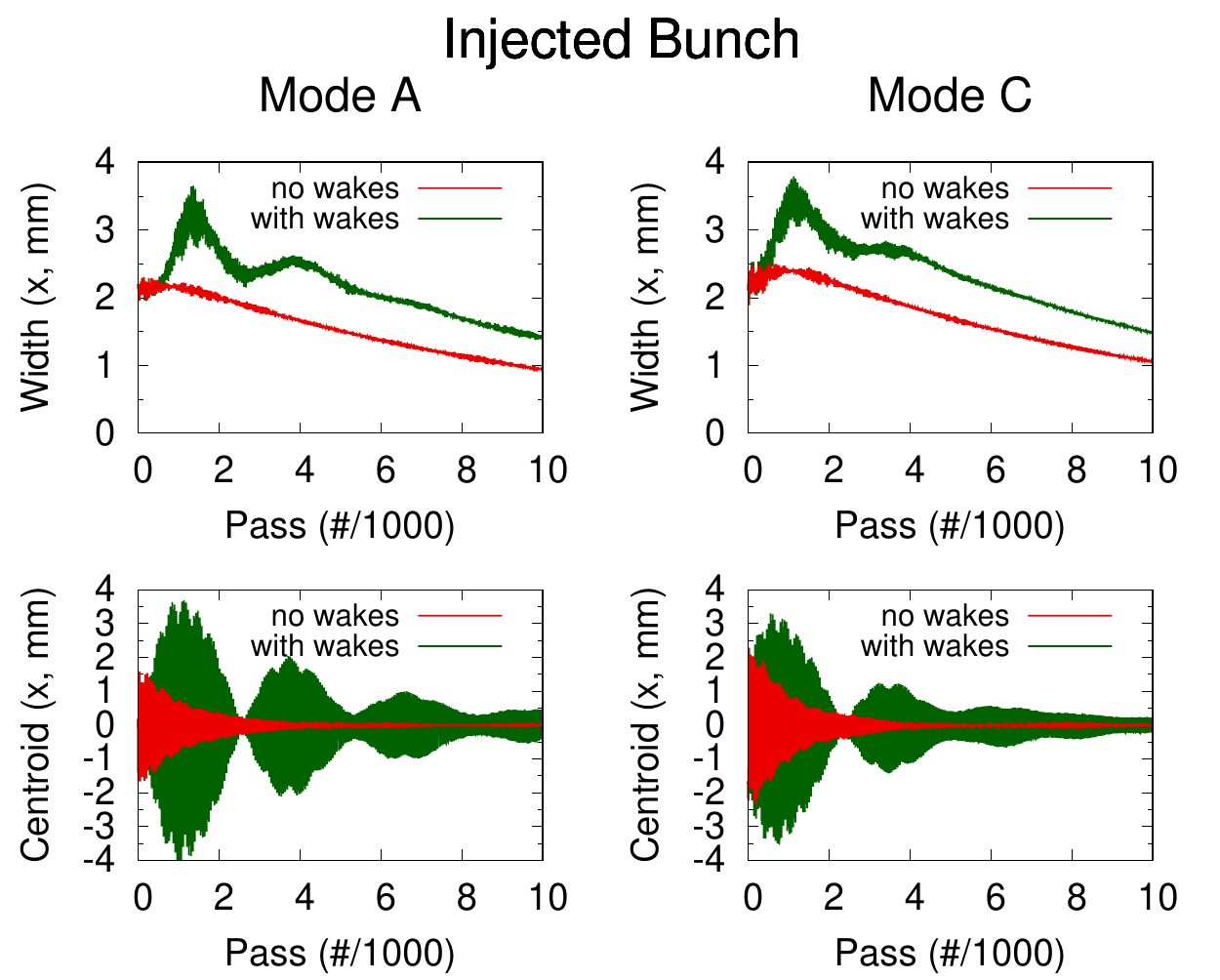}
 \caption{\label{fig:cases_inject}
 Comparison of the evolution of the size (top) and centroid motion (bottom) of the injection bunch 
 for the cases without wake fields (red) and with nominal wake fields (green). 
 The left figures show results for the 3DK settings referred to as Mode A (`balanced'),
 and right figures show Mode C, which is tuned for better performance in the presence 
 of wake fields.  The corresponding injection efficiencies for the two cases are 
 83.6\% and 93.1\%.}
\end{figure}

The 3DK `balanced' configuration (Mode A) was chosen by optimizing for injection
efficiency, but without considering the impact of wakes.  Less than 
$1$\% beam losses are predicted in this case.  
However, when the expected
wake fields are included, the injection efficiency drops to $83.6$\%.
As shown in Fig.~\ref{fig:cases_inject}, most of the losses occur between $0.5$ ms and $1$ ms 
after injection, roughly between $1000$ and $2000$ passes around the ring. 

An important aspect of the 3DK scheme is that it is flexible enough to allow for additional
adjustments to accommodate the impact of wake fields.  An alternate 
set of all three kicker magnet amplitudes, referred to as `Mode B', was found that
reduced the displacement of the trajectory of the stored beam by $20$\%,
while increasing losses by a modest amount when wake fields are neglected.  
These additional losses occur in the first 50 passes around the ring.
The capture efficiency remains above $99$\%.
For the nominal wake fields the capture efficiency is $93.2$\%, a significant improvement 
compared to $83.6$\% for Mode A.  In none of these cases are there losses in the stored bunch.

Mode C is similar to Mode B except that the septum field is weaker and the horizontal beta function 
at the end of the septum is lower than the matched value so as to reduce transfer-line
losses on the injection septum.  These adjustments also reduce the injected beam transient slightly, 
though when wake fields are taken into account the capture efficiency 
is comparable to that of Mode B.  The performance of Modes B and C are much more similar to each other than they are 
to Mode A, and comparisons will focus on the differences between Mode A and Mode C.

Figure \ref{fig:cases_inject} shows
the evolution of the size and horizontal offset of the injected bunch when 
short-range wake fields are either included or ignored.  Results for Mode A are 
shown on the left-side plots, and for Mode C on the right-side plots.   
In Mode A, the width of the injected bunch steadily decays
without wake fields, while with the nominal wakes the width of the bunch 
grows until it peaks after $1800$ turns around the ring
with an rms of $3.5$ mm, up from $2.2$ mm.  The centroid motion rapidly damps from $1.5$ mm 
without wakes, but with the nominal wakes it increases to $3.5$ mm amplitude, followed by 
envelope oscillations which slowly decay over thousands of passes around the ring.
For Mode C, there is similar behavior in the width of the injected bunch but 
the amplitude of the centroid motion starts out at $2$ mm and only grows to $3.0$ mm.
The centroid motion again has continued envelope oscillations in the presence of wakes,
but at a lower amplitude than for Mode A.  

For the nominal wake fields, the goal of at least $95$\% injection efficiency is not met
even for Mode C.
Thus, other strategies to mitigate the impact of short-range wake fields have been examined.
The nominal chromaticity of $1$ in both planes can easily be increased and will drive 
more rapid phase decoherence of the bunches.
Both horizontal and vertical chromaticities contribute to this effect at high amplitudes.
A small temporal offset of the injected bunch relative to the stored bunch
directly reduces the wake field forces experienced by the injected electrons.
%These strategies, together with further adjustment of the 3DK settings,
%should lead to even more robustness against short-range wake field effects.

Increasing the chromaticity from $1$ to $1.2$ increases the capture
efficiency to $93.5$\% for Mode A and $97.7$\% for Mode C.
Similarly, keeping the chromaticity at unity and delaying the injected bunch
by $100$ ps relative to the stored bunch leads to an capture efficiency of
$98.5$\% for Mode A nd $98.9$\% for Mode C.  
Applying both changes brings the capture efficiency above $99$\% in both modes.

\begin{figure}[htb]
\centering
	\includegraphics[width=0.48\textwidth]{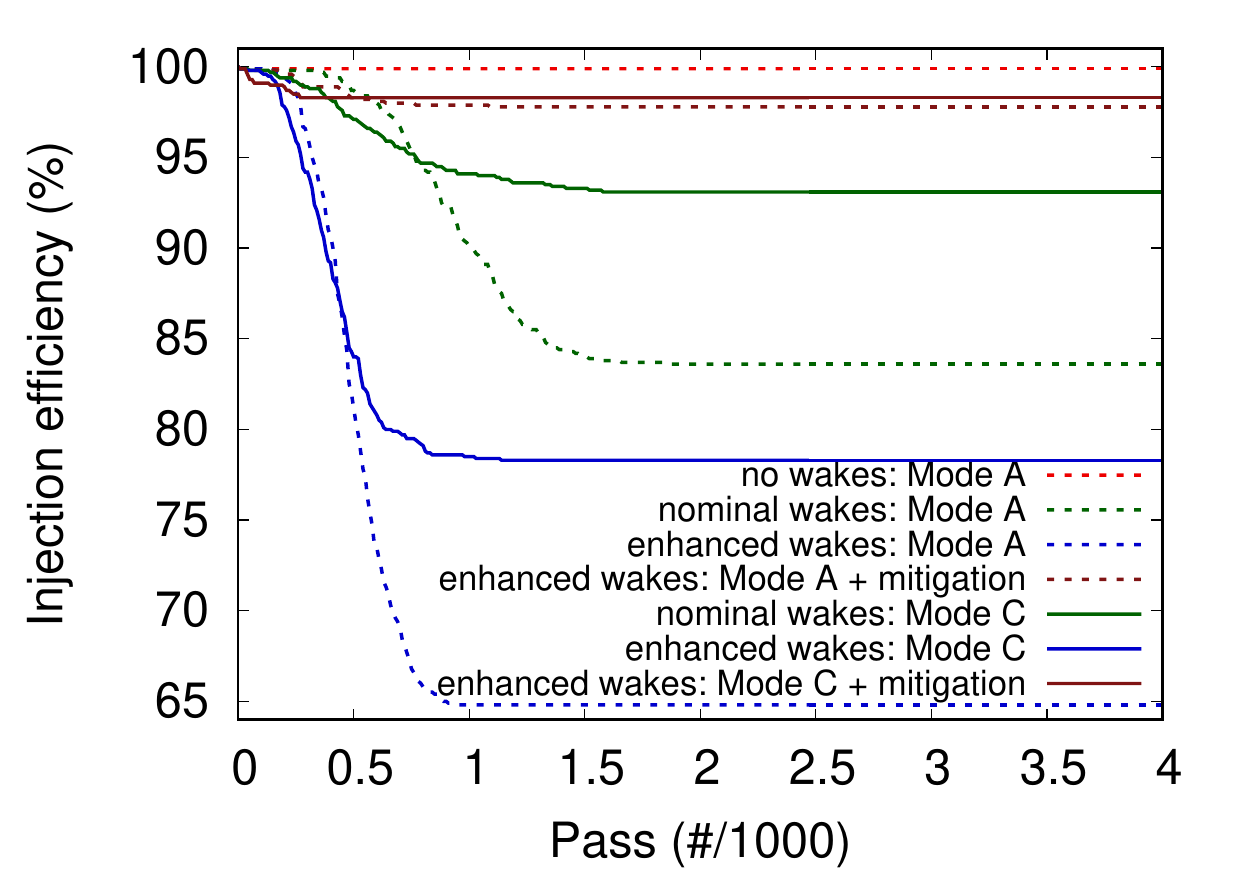}
 \caption{\label{fig:wake_inj_efficiency}
 Capture efficiency as a function of number of passes around the ring after injection,
 comparing Mode A, where the kicker settings are optimized without including wake field effects,
 to Mode C, which uses an alternative set of kicker strengths which is more tolerant 
 to wake fields and also changes the injection beta function. Results are shown with 
 and without wake fields, for enhanced wake fields, and for additional mitigations.}
\end{figure}

To exercise a measure of caution, we considered the possibility that wake fields would be
twice as high as expected (``enhanced wakes'')
%, such that the TMCI  threshold would be reduced to $3$ nC.
In this case, the capture  efficiency drops to $64.8$\% for Mode A and $78.3$\% for Mode C without additional mitigations.
The evolution of the capture efficiency as a function of number of passes after
injection is shown in Fig.~\ref{fig:wake_inj_efficiency} for both Mode A and Mode C.
The wake field levels are varied among no wakes, nominal wakes and enhanced wakes.
In addition, some examples include using a combined mitigation of setting both chromaticities 
to $1.2$ and shifting the injected bunch by $100$ ps.  For the enhanced wakes, this 
brings the capture efficiency to $97.8$\% for Mode A and $98.3$\% for Mode C.
Creating a temporal offset between the stored and injected bunches 
is particularly effective at reducing the impact of wake fields.  For Mode C with the above mitigations, 
the capture efficiency is 97.2\% when the 
strength of the wake fields is multiplied by a factor of three.  The corresponding
efficiency for Mode A is $95.9$\%.

\section{Multi-bunch stability and interference with TFB system\label{sec:TFB}}
The transverse bunch-by-bunch feedback system (TFB) will detect the transient of the stored beam and apply 
kicks to damp it.  Unfortunately, those kicks will anti-damp the injected particles since the 
stored beam and injected particles are out of phase.  The injected particles 
carry relatively little charge and will offset only slightly the center-of-charge that the 
feedback system detects.  To prevent the loss of injected particles as the TFB damps the
injection transient, the TFB will be masked for those buckets into which charge is injected.
A single injection shot delivers charge into $4$ buckets, spaced $4$ buckets apart,
for example, buckets $10$, $14$, $18$, and $22$.  Masking the buckets significantly diminishes the 
ability of the TFB to damp multibunch instabilities and so we investigate here the injection 
transient and TFB interaction using a tracking simulation.

\begin{figure}[!t]
\centering
	\includegraphics[width=0.48\textwidth]{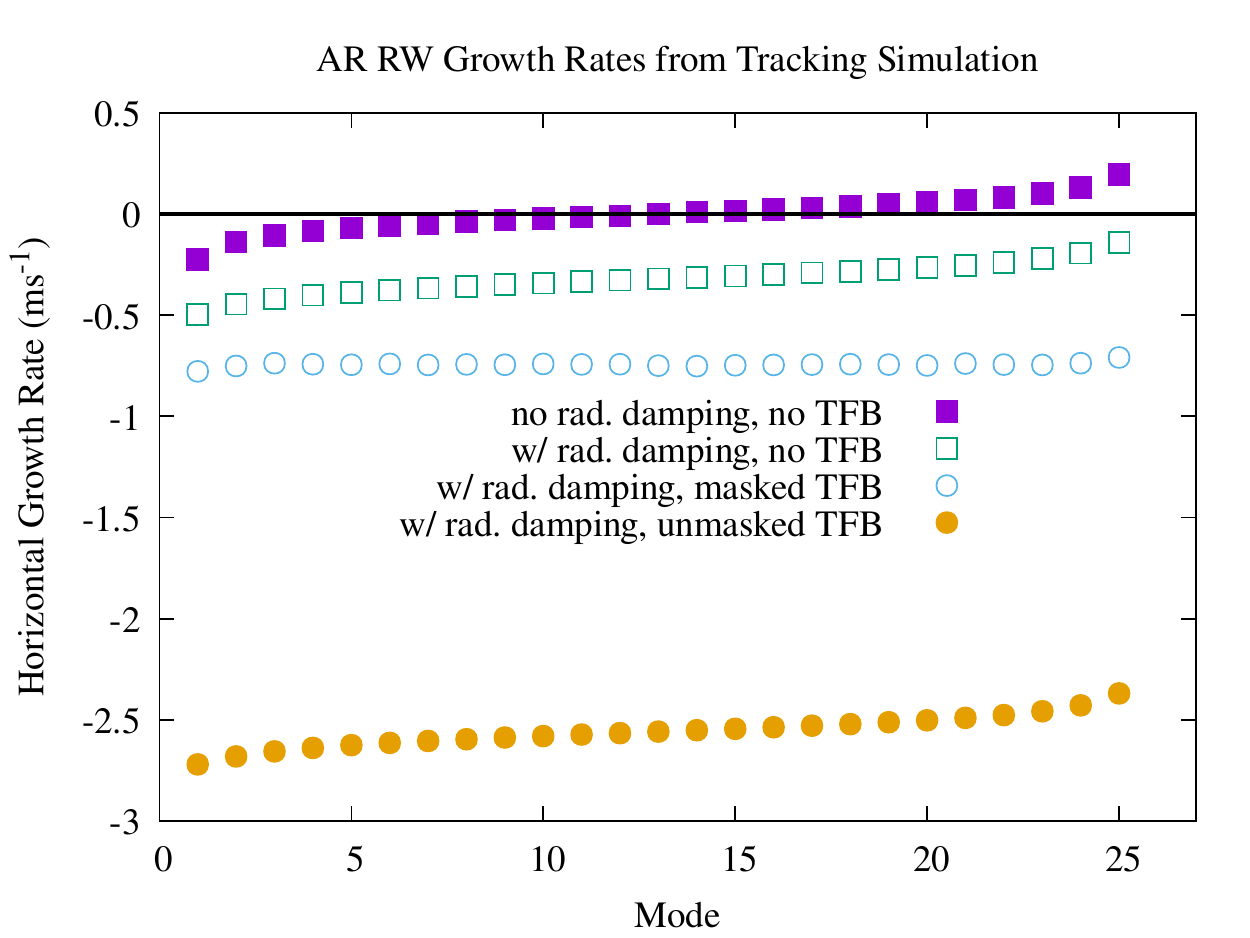}\\
	\includegraphics[width=0.48\textwidth]{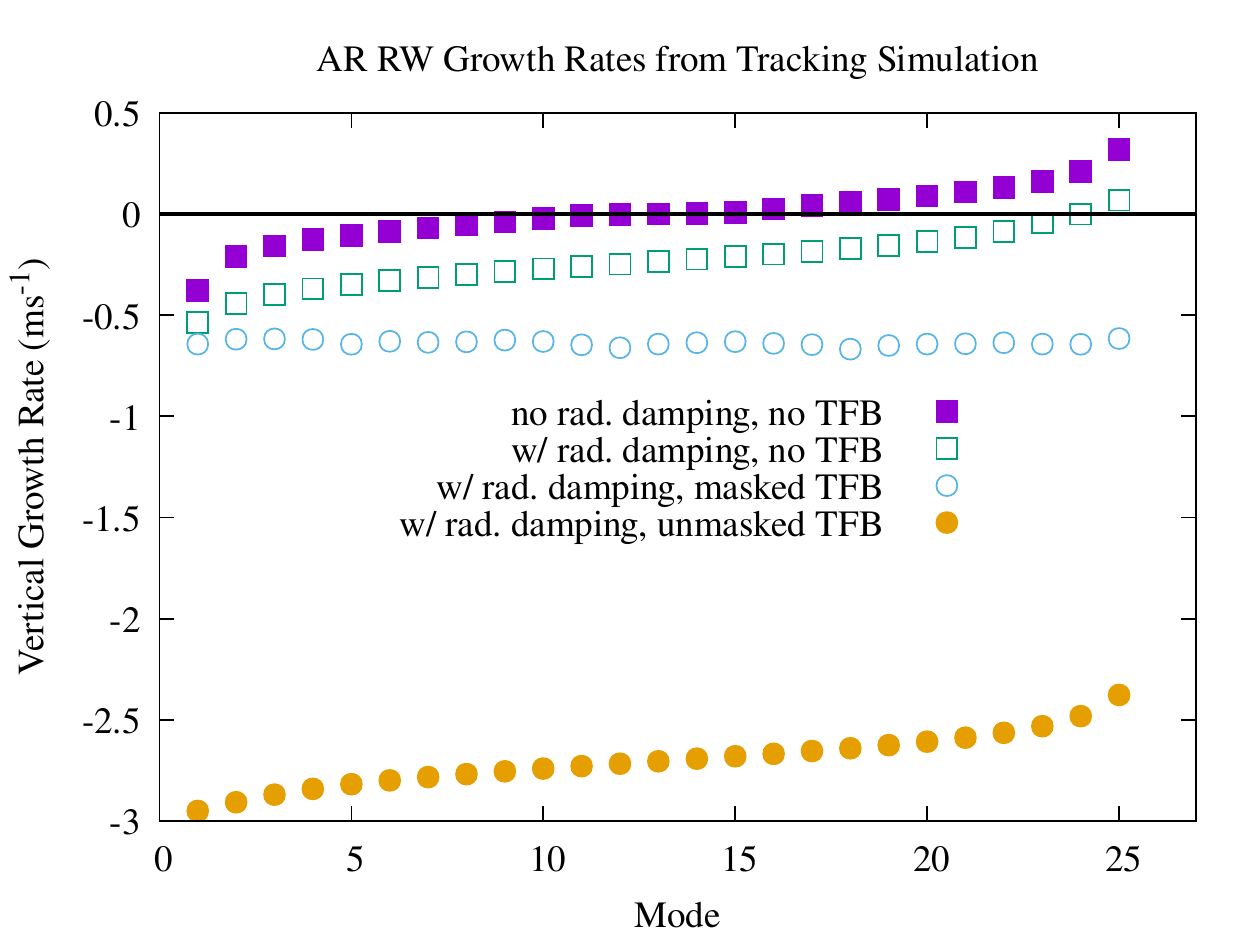}\\
 \caption{\label{fig:multibunch-rw-rates}
 Per-mode growth rates of a $25$ bunch train in the accumulator ring determined by multi-bunch tracking with resistive
 wall wake fields.  During steady-state, all RW modes are safely damped with a total growth rate of about $-2.5$ ms$^{-1}$.  During the injection transient the effectiveness of the TFB is significantly diminished by the necessity of masking the $4$ 
 buckets into which charge is injected, though all modes remain damped.}
\end{figure}

Under nominal conditions, the growth rates of multibunch impedance instabilities in the AR are 
less than that of the radiation damping and multibunch feedback system.
The accumulator-ring HOM wakes have been evaluated with theory and tracking
simulations.  The strongest cavity HOM growth rate is $0.03$ ms$^{-1}$, and
the radiation damping rate is $0.16$ ms$^{-1}$.
The growth rates of the resistive wall modes are shown in Fig.~\ref{fig:multibunch-rw-rates}.  
Except for the two highest order modes, all are below the radiation damping time,
and the highest order modes are well-within the capabilities of the TFB.

Following an injection cycle the stored beam is left with a
large transverse offset and the $4$ buckets into which charge is injected are masked
from the multi-bunch feedback.  Since the resistive wall modes are near or exceed the radiation 
damping rate, a tracking simulation
is used to evaluate their stability during the injection transient when the effectiveness of the TFB is 
diminished.

The simulation models the long-range resistive wall wakes at any position in the ring using a fit of $15$ pseudo-modes to 
the $1/\sqrt{t}$ dependence of the long-range wake fields and follows closely the technique described in      \cite{Wang:IPAC2017-WEPIK110}.  The long-range resistive wall wake is,
\begin{equation}
W\left(t\right)=\frac{\sqrt{c}}{\pi b^3}\sqrt{\frac{Z_0}{\pi\sigma}}\frac{L}{\sqrt{t}},
\label{eqn:Wt}
\end{equation}
where $b$ is the chamber dimension and $\sigma$ its conductivity;  $c$ and $Z_0$ are the
speed of light and impedance of free space;  $L$ is the length of the chamber and $t$ is time  since the passage of the source particle\cite{Chao:1993zn}.
The basis of pseudo-modes is
\begin{equation}
W_i\left(t\right)=\frac{A_i^2}{S_i^2}\exp\left(-\frac{d_i^2}{S_i^2}t\right),
\end{equation}
where $A_i$, $S_i$, and $d_i$ are fit parameters. Here, $S_i$ is redundant but including it as a fit parameter
has been found to help the fitter converge consistently and quickly with simply default parameters.
The fitter is {\tt Mathematica}'s {\tt NonlinearModelFit} \cite{Mathematica}.  The log of the basis is fit to 
$\log{\left(1/\sqrt{t}\right)}$ sampled with a logarithmic distribution of $t$.
The residuals of the fit are less than $0.1\%$ from $1$ ns to $44,000$ turns, and reliably decay to zero 
beyond the ends of the fit.  The same basis is used for all chambers, but is scaled according
to the local chamber material and dimension shown in Eq.~(\ref{eqn:Wt}).  Resistive wall wakes are 
represented at $216$ locations in the ring.

For each horizontal mode $n$, each bunch $i$  in the train is seeded with an appropriate offset,
\begin{align}
    x_i =& \sqrt{J\beta_x}\cos\left(\frac{\pi}{2}+\frac{2\pi n i}{N_{bunches}}\right)\label{eqn:seedx}\\
    x'_i =& \sqrt\frac{J}{\beta_x}\Big(
    \alpha_x\cos\left(\frac{\pi}{2}+\frac{2\pi n i}{N_{bunches}}\right)\nonumber\\
    &+\sin\left(\frac{\pi}{2}+\frac{2\pi n i}{N_{bunches}}\right)
    \Big)\label{eqn:seedpx}.
\end{align}
and tracked for $5000$ turns.

The multi-bunch tracking simulation is developed using {\tt Bmad} \cite{Sagan:Bmad2006}, which keeps track of the pseudo-modes
as they build up and influence the beam turn-by-turn.
Radiation damping is implemented by a per-turn decrement to the action of each macro-particle.
The multibunch feedback is simulated using a pickup, FIR filter, and kicker.

Each turn the normal mode coordinates of each bunch are recorded and a Fourier transform is taken
along the bunch train.  For the seeded mode, the growth rate is obtained from the fit of an exponential to the 
the height of the spectral peak versus turn number.
This is done similarly for the vertical modes.
The growth rates of each of the $25$ modes are shown in 
Fig.~\ref{fig:multibunch-rw-rates}.
From Fig.~\ref{fig:multibunch-rw-rates} we conclude that while masking will significantly reduce the 
damping obtained from the multibunch feedback system, all resistive wall will modes remain stable 
by comfortable margin during the transient.

Following an injection cycle, the bunch train is offset with a large uniform transient.
However, the study just described used small amplitudes and particular offsets given by Eqs.~(\ref{eqn:seedx}) and (\ref{eqn:seedpx}).  To simulate absolute stability following an injection cycle, the beam is 
given a uniform $5$ mm horizontal offset and tracked for $5000$ turns.  
Without masking,  after $1702$ turns the bunch oscillating with the largest amplitude has an action that
is $50$\% of the initial condition.  With $4$ bunches masked, $50$\% is achieved after $3266$ turns.
All $15$ RW pseudo-modes are reliably decaying in strength throughout the simulation.

\section{Conclusion}
The three-dipole kicker injection scheme is well-fit to the particular constraints and allowances 
of the ALS-U accumulator ring.  
It reliably injects the relatively large beam coming out of the booster with nearly $100\%$ efficiency using
conventional technology.  It meets the accumulator ring's tight space constraints by distributing 
the injection kickers across multiple straights.  It fits the large beam coming in from the booster
into the relatively small vacuum chamber by allowing a transient on the stored beam so as to shrink the
transient of the injected beam.

The injection transient has consequences for long- and short-range
wake fields, but several techniques have been prepared to effectively mitigate these consequences.
The scheme is robust in the presence of lattice and pulsed element imperfections.
Both the wake field issues and imperfection tolerances benefit from the ability of the scheme
to tune the amplitudes of the stored and injected transients.

In the appendix, the merits of three alternative injection schemes are compared to those of the 3DK scheme.

\appendix

\section{Alternate Injection Options}
\label{App:A}

\subsection{Nonlinear Kicker}

%\begin{figure}[b]
%\centering
%\includegraphics[width=0.48\textwidth]{Hist_Inj_eff_Dir_15764774_postLoco.pdf}
%\caption{Histogram of injection efficiency evaluated for 95 post-LOCO lattices.}
%\label{fig:NLK_Inj_Hist}
%\end{figure}

\begin{figure}[!b]
\centering
	\includegraphics[width=0.4\textwidth]{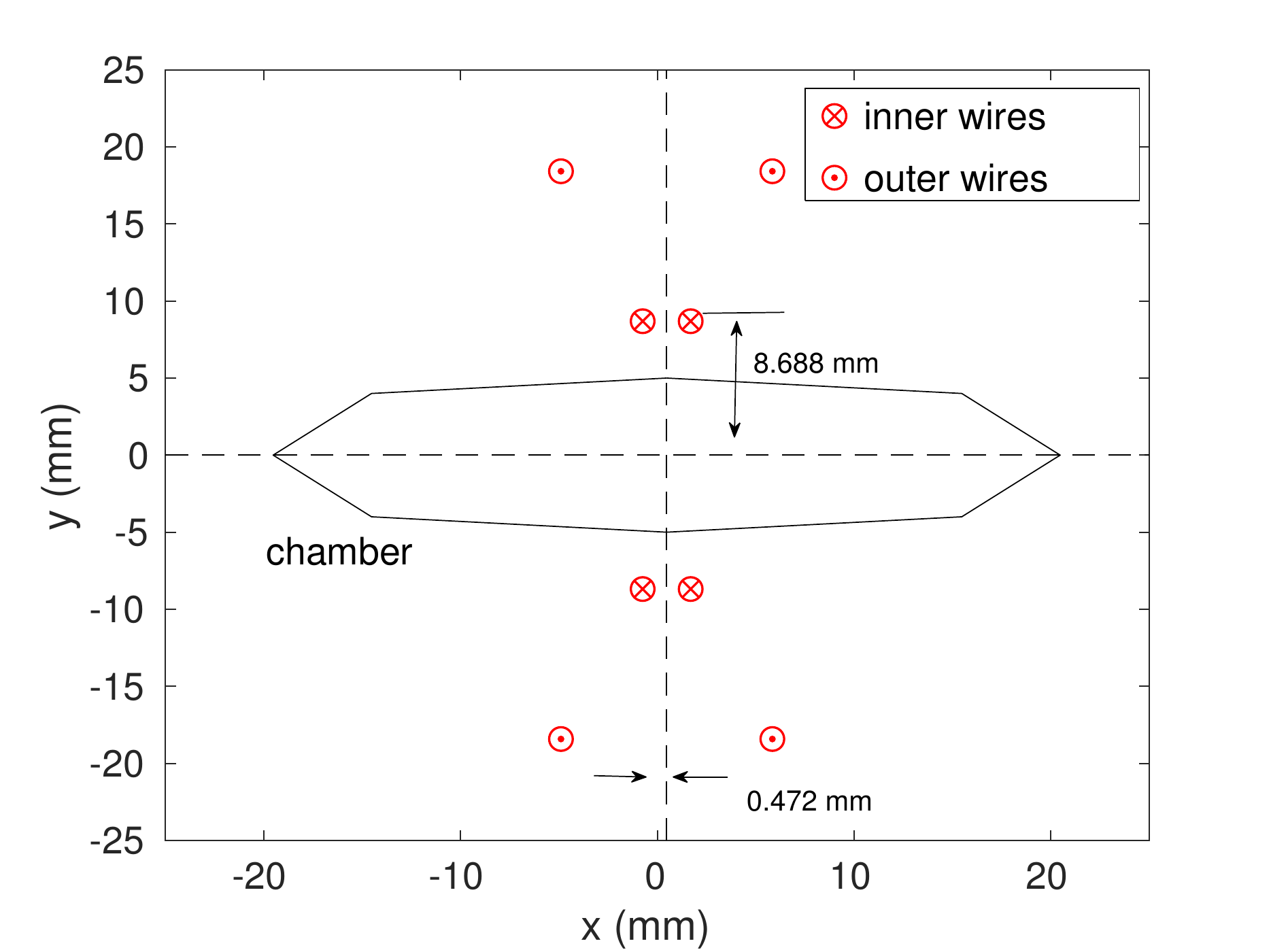}\\
	\includegraphics[width=0.41\textwidth]{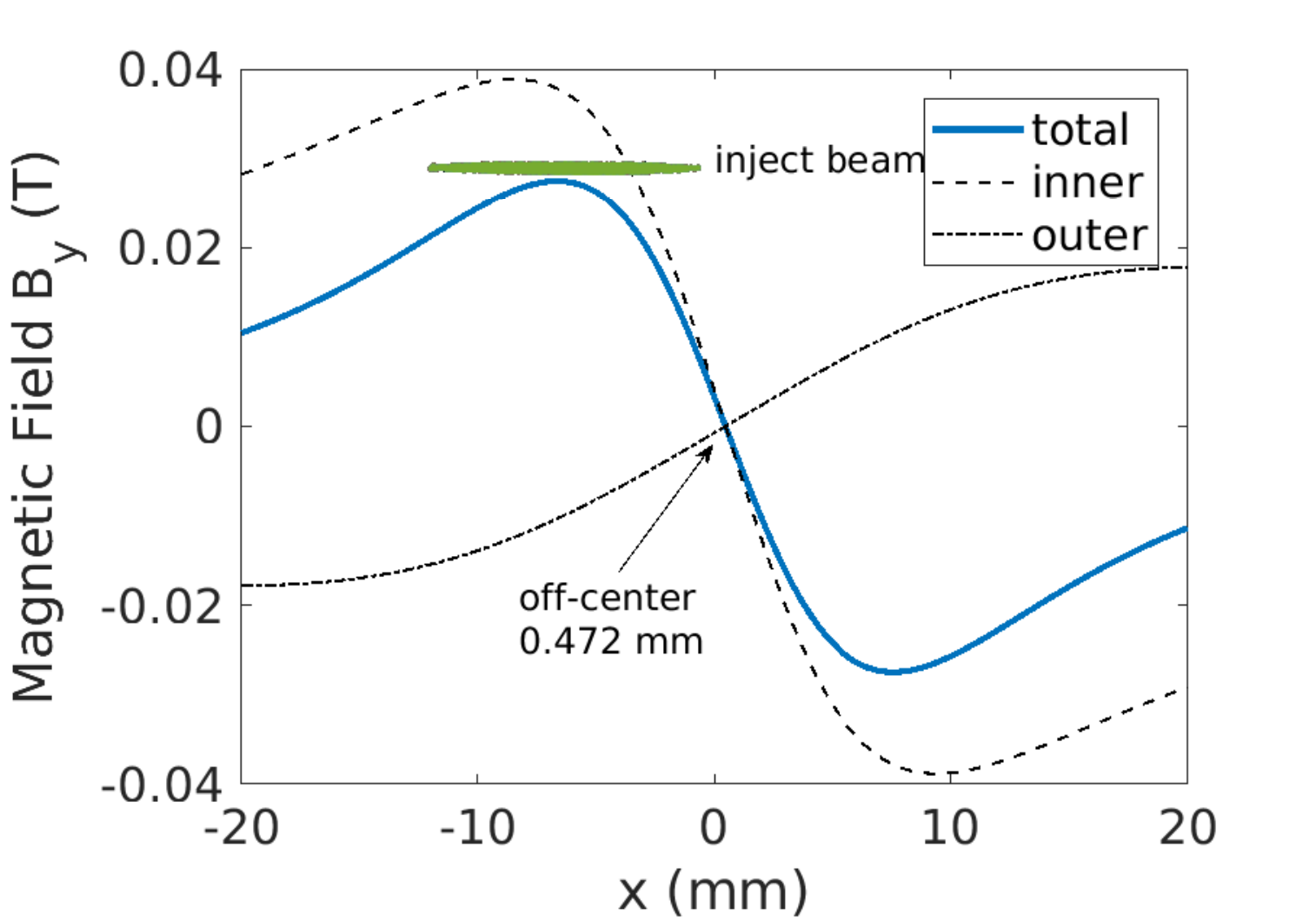}
 \caption{\label{fig:NLK_wire_field}
 NLK-design concept. Top: main conductors'  transverse positions (the return conductors are not shown). Bottom:  magnetic-field profile in the mid plane.    }
\end{figure}

Unlike conventional dipole-kicker magnets, a  nonlinear kicker (NLK) generates a transverse nonlinear  magnetic-field profile with a maximum  located off-axis at the injected-beam arrival point and zero in the center. The injected beam is kicked while the stored beam at the center is minimally perturbed.  In addition, a single NLK device is sufficient for injection, thus easing issues of space and synchronization. Because of these reasons,  several light source facilities have adopted NLK in alternative to more conventional closed-orbit bump schemes \cite{PhysRevSTAB.15.050705, Rast:2011zz, Liu:2016rcc, ALEXANDRE2021164739}.

Although injection perturbations to the stored beam  are not a major concern in the  AR (the stored beam does not serve user experiments), NLK injection represents an attractive option because of its compactness and simplicity, while a relaxed tolerance  to stored-beam perturbation can be exploited to optimize the design in ways that are not suitable for light-source applications.  

The NLK design configuration we considered consisted  of $8$ main conductors symmetrically positioned around the vacuum chamber in the direction of the beam as shown in Fig.~\ref{fig:NLK_wire_field}.  The NLK was located in the same lattice position as the main kicker in the 3DK scheme.  The conductor placement was optimized using MOGA  methods \cite{996017} in start-to-end macroparticle injection simulations; for a detailed description of the optimization setup and results see Ref.~\cite{PhysRevAccelBeams.23.010702}. With an optimized NLK design, injection tracking studies in the presence of lattice errors and including orbit/optics correction showed injection efficiency above 95\%  (not including transverse wakefield effects).

While these simulations  were encouraging, an NLK-solution for the AR injection was eventually abandoned after the R\&D effort initiated at ALS~\cite{PhysRevAccelBeams.23.010702} encountered technical difficulty with  the  prototyping of the  ceramic vacuum-chamber.

\subsection{Conventional closed orbit bump 4DK injection in one sector}
The existing ALS utilizes a conventional $4$-kicker orbit-bump injection scheme with all four kickers in the same straight section.   A similar scheme was explored for the AR
with all $4$ injection kickers placed in Sector $2$.  Potential advantages of this scheme are:
\begin{enumerate*}[1) ]
\item it is conventional,
\item residual oscillations of the stored beam following the injection cycle are eliminated,
\item it eliminates the need to carefully control the phase advance between the pre-kicker and main injection kicker.
\end{enumerate*}

\begin{figure}[hb]
\centering
\includegraphics[width=0.48\textwidth]{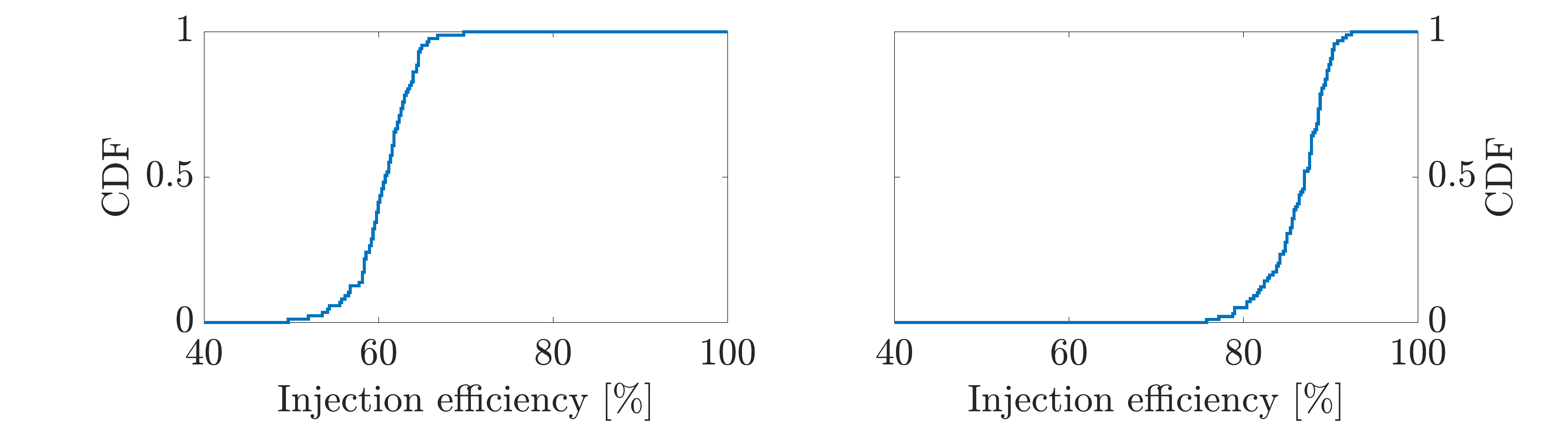}
 \caption{Injection efficiency for the conventional 4DK scheme evaluated over a population of 100  lattice-error realizations including orbit and optics corrections.  Results are for the septum positioned at \SI{8}{\mm} (left) and  \SI{13}{\mm} (right) from the vacuum chamber center. Tracking simulations were done with $10$k particles over 10k machine turns.}
 \label{fig:4DK}
\end{figure}

The straights in the AR are $0.624$ m shorter than those in the ALS.
This shorter space drives up the the strength required from the dipole kickers.  Additionally, because the orbit bump is   closed for the stored beam, the injected particles oscillate with larger amplitude.  To avoid loss of injected particles, the injection septum aperture would need to be increased to the nominal vacuum chamber aperture of about $14$ mm.
This complicates engineering in the injection straight and further increases the necessary dipole kick 
to about $10$ mrad.
Such a large dipole kick requires some combination of kicker R\&D, a multi-turn ramp, or placing the $4$ kickers on
a single power supply string.  A multi-turn ramp would constrain the machine tune, thus negating one potential 
advantage of this scheme.  Placing the $4$ kickers on a single
string would make it difficult to trade off between stored beam and injected beam oscillation amplitudes, which 
is a disadvantage compared to the 3DK scheme. 
An additional drawback is that the septa would need to be moved upstream (toward the center of the straight) 
to make room for the 4 kickers.  This has the consequence of forcing a stronger bending requirement 
in the thick septum.

Even with a $14$ mm septum aperture, a conventional $4$-kicker design delivers less than $90$\% 
injection efficiency, as shown in Fig.~\ref{fig:4DK}. In contrast, the 3DK scheme approaches $100$\% injection 
efficiency.  The 3DK scheme also compares 
well in requiring only an $8$ mm septum aperture, and its kicker technology requirements are modest: 
three $0.6$ m long ferrite-loaded dipole kickers delivering $\sim1$ mrad kick.

\subsection{Two dipole kicker (2DK)}
It is possible to implement an injection scheme similar to 3DK while using only two dipole kickers.  Coined ``2DK''
this scheme saves costs by requiring one less kicker.  With 3DK, the two pre-kickers allow adjustment of
the position and angle of the stored beam at the main injection kicker.  With 2DK, only one
pre-kicker is employed and its location in the storage ring is chosen to have a horizontal phase advance from
the pre-kicker to the main injection kicker such that the stored beam has a near-zero crossing at the main injection
kicker.  The pre-kicker is placed in sector $7$, which is six sectors upstream of the main injection kicker.  

2DK is competitive as far as injection efficiency goes, but the lack of flexibility limits the ability 
to accommodate difficult-to-predict consequences of imperfections and wake fields.  It also constrains the
ring tune, as the pre-kicker to main injection kicker phase advance must be precisely set.  Generally, the phase
advance along a single straight is too little to allow for significant adjustment.  It was judged that the versatility
of the 3DK scheme justified the cost.

\bibliography{main}

\begin{thebibliography}{15}%
\makeatletter
\providecommand \@ifxundefined [1]{%
 \@ifx{#1\undefined}
}%
\providecommand \@ifnum [1]{%
 \ifnum #1\expandafter \@firstoftwo
 \else \expandafter \@secondoftwo
 \fi
}%
\providecommand \@ifx [1]{%
 \ifx #1\expandafter \@firstoftwo
 \else \expandafter \@secondoftwo
 \fi
}%
\providecommand \natexlab [1]{#1}%
\providecommand \enquote  [1]{``#1''}%
\providecommand \bibnamefont  [1]{#1}%
\providecommand \bibfnamefont [1]{#1}%
\providecommand \citenamefont [1]{#1}%
\providecommand \href@noop [0]{\@secondoftwo}%
\providecommand \href [0]{\begingroup \@sanitize@url \@href}%
\providecommand \@href[1]{\@@startlink{#1}\@@href}%
\providecommand \@@href[1]{\endgroup#1\@@endlink}%
\providecommand \@sanitize@url [0]{\catcode `\\12\catcode `\$12\catcode
  `\&12\catcode `\#12\catcode `\^12\catcode `\_12\catcode `\%12\relax}%
\providecommand \@@startlink[1]{}%
\providecommand \@@endlink[0]{}%
\providecommand \url  [0]{\begingroup\@sanitize@url \@url }%
\providecommand \@url [1]{\endgroup\@href {#1}{\urlprefix }}%
\providecommand \urlprefix  [0]{URL }%
\providecommand \Eprint [0]{\href }%
\providecommand \doibase [0]{http://dx.doi.org/}%
\providecommand \selectlanguage [0]{\@gobble}%
\providecommand \bibinfo  [0]{\@secondoftwo}%
\providecommand \bibfield  [0]{\@secondoftwo}%
\providecommand \translation [1]{[#1]}%
\providecommand \BibitemOpen [0]{}%
\providecommand \bibitemStop [0]{}%
\providecommand \bibitemNoStop [0]{.\EOS\space}%
\providecommand \EOS [0]{\spacefactor3000\relax}%
\providecommand \BibitemShut  [1]{\csname bibitem#1\endcsname}%
\let\auto@bib@innerbib\@empty
%</preamble>
\bibitem [{\citenamefont {Steier}\ \emph {et~al.}(2018)\citenamefont {Steier}
  \emph {et~al.}}]{Steier:IPAC2018-THPMF036}%
  \BibitemOpen
  \bibfield  {author} {\bibinfo {author} {\bibfnamefont {C.}~\bibnamefont
  {Steier}} \emph {et~al.},\ }in\ \href {\doibase
  doi:10.18429/JACoW-IPAC2018-THPMF036} {\emph {\bibinfo {booktitle} {Proc. 9th
  International Particle Accelerator Conference (IPAC'18), Vancouver, BC,
  Canada, April 29-May 4, 2018}}},\ \bibinfo {series and number} {\bibinfo
  {series} {International Particle Accelerator Conference}\ No.~\bibinfo
  {number} {9}}\ (\bibinfo  {publisher} {JACoW Publishing},\ \bibinfo {address}
  {Geneva, Switzerland},\ \bibinfo {year} {2018})\ pp.\ \bibinfo {pages}
  {4134--4137},\ \bibinfo {note}
  {https://doi.org/10.18429/JACoW-IPAC2018-THPMF036}\BibitemShut {NoStop}%
\bibitem [{\citenamefont {Leemann}(2012{\natexlab{a}})}]{LEEMANN2012117}%
  \BibitemOpen
  \bibfield  {author} {\bibinfo {author} {\bibfnamefont {S.}~\bibnamefont
  {Leemann}},\ }\href {\doibase https://doi.org/10.1016/j.nima.2012.07.023}
  {\bibfield  {journal} {\bibinfo  {journal} {Nuclear Instruments and Methods
  in Physics Research Section A: Accelerators, Spectrometers, Detectors and
  Associated Equipment}\ }\textbf {\bibinfo {volume} {693}},\ \bibinfo {pages}
  {117} (\bibinfo {year} {2012}{\natexlab{a}})}\BibitemShut {NoStop}%
\bibitem [{\citenamefont {Streun}(2005)}]{streun-inj-mismatch}%
  \BibitemOpen
  \bibfield  {author} {\bibinfo {author} {\bibfnamefont {A.}~\bibnamefont
  {Streun}},\ }\href {https://ados.web.psi.ch/slsnotes/tmeta020193.pdf} {\emph
  {\bibinfo {title} {SLS booster-to-ring transferline optics for optimum
  injection efficiency}}},\ \bibinfo {type} {Tech. Rep.}\ \bibinfo {number}
  {SLS-TME-TA-2002-0193}\ (\bibinfo  {institution} {Paul Scherrer Institut},\
  \bibinfo {address} {Villigen PSI, Switzerland},\ \bibinfo {year}
  {2005})\BibitemShut {NoStop}%
\bibitem [{\citenamefont {Sun}\ \emph {et~al.}(2019)\citenamefont {Sun} \emph
  {et~al.}}]{Sun:IPAC2019-WEPGW111}%
  \BibitemOpen
  \bibfield  {author} {\bibinfo {author} {\bibfnamefont {C.}~\bibnamefont
  {Sun}} \emph {et~al.},\ }in\ \href {\doibase
  doi:10.18429/JACoW-IPAC2019-WEPGW111} {\emph {\bibinfo {booktitle} {Proc.
  10th International Particle Accelerator Conference (IPAC'19), Melbourne,
  Australia, 19-24 May 2019}}},\ \bibinfo {series and number} {\bibinfo
  {series} {International Particle Accelerator Conference}\ No.~\bibinfo
  {number} {10}}\ (\bibinfo  {publisher} {JACoW Publishing},\ \bibinfo
  {address} {Geneva, Switzerland},\ \bibinfo {year} {2019})\ pp.\ \bibinfo
  {pages} {2756--2759},\ \bibinfo {note}
  {https://doi.org/10.18429/JACoW-IPAC2019-WEPGW111}\BibitemShut {NoStop}%
\bibitem [{\citenamefont {Deb}\ \emph {et~al.}(2002)\citenamefont {Deb},
  \citenamefont {Pratap}, \citenamefont {Agarwal},\ and\ \citenamefont
  {Meyarivan}}]{996017}%
  \BibitemOpen
  \bibfield  {author} {\bibinfo {author} {\bibfnamefont {K.}~\bibnamefont
  {Deb}}, \bibinfo {author} {\bibfnamefont {A.}~\bibnamefont {Pratap}},
  \bibinfo {author} {\bibfnamefont {S.}~\bibnamefont {Agarwal}}, \ and\
  \bibinfo {author} {\bibfnamefont {T.}~\bibnamefont {Meyarivan}},\ }\href
  {\doibase 10.1109/4235.996017} {\bibfield  {journal} {\bibinfo  {journal}
  {IEEE Transactions on Evolutionary Computation}\ }\textbf {\bibinfo {volume}
  {6}},\ \bibinfo {pages} {182} (\bibinfo {year} {2002})}\BibitemShut {NoStop}%
\bibitem [{\citenamefont {Hellert}\ \emph {et~al.}(2019)\citenamefont
  {Hellert}, \citenamefont {Amstutz}, \citenamefont {Steier},\ and\
  \citenamefont {Venturini}}]{PhysRevAccelBeams.22.100702}%
  \BibitemOpen
  \bibfield  {author} {\bibinfo {author} {\bibfnamefont {T.}~\bibnamefont
  {Hellert}}, \bibinfo {author} {\bibfnamefont {P.}~\bibnamefont {Amstutz}},
  \bibinfo {author} {\bibfnamefont {C.}~\bibnamefont {Steier}}, \ and\ \bibinfo
  {author} {\bibfnamefont {M.}~\bibnamefont {Venturini}},\ }\href {\doibase
  10.1103/PhysRevAccelBeams.22.100702} {\bibfield  {journal} {\bibinfo
  {journal} {Phys. Rev. Accel. Beams}\ }\textbf {\bibinfo {volume} {22}},\
  \bibinfo {pages} {100702} (\bibinfo {year} {2019})}\BibitemShut {NoStop}%
\bibitem [{\citenamefont {Wang}\ \emph {et~al.}(2017)\citenamefont {Wang},
  \citenamefont {Billing}, \citenamefont {Poprocki}, \citenamefont {Rubin},\
  and\ \citenamefont {Sagan}}]{Wang:IPAC2017-WEPIK110}%
  \BibitemOpen
  \bibfield  {author} {\bibinfo {author} {\bibfnamefont {S.}~\bibnamefont
  {Wang}}, \bibinfo {author} {\bibfnamefont {M.}~\bibnamefont {Billing}},
  \bibinfo {author} {\bibfnamefont {S.}~\bibnamefont {Poprocki}}, \bibinfo
  {author} {\bibfnamefont {D.~L.}\ \bibnamefont {Rubin}}, \ and\ \bibinfo
  {author} {\bibfnamefont {D.}~\bibnamefont {Sagan}},\ }in\ \href {\doibase
  https://doi.org/10.18429/JACoW-IPAC2017-WEPIK110} {\emph {\bibinfo
  {booktitle} {Proc. of International Particle Accelerator Conference
  (IPAC'17), Copenhagen, Denmark, 14-19 May, 2017}}},\ \bibinfo {series and
  number} {\bibinfo {series} {International Particle Accelerator Conference}\
  No.~\bibinfo {number} {8}}\ (\bibinfo  {publisher} {JACoW},\ \bibinfo
  {address} {Geneva, Switzerland},\ \bibinfo {year} {2017})\ pp.\ \bibinfo
  {pages} {3204--3207},\ \bibinfo {note}
  {https://doi.org/10.18429/JACoW-IPAC2017-WEPIK110}\BibitemShut {NoStop}%
\bibitem [{\citenamefont {Chao}(1993)}]{Chao:1993zn}%
  \BibitemOpen
  \bibfield  {author} {\bibinfo {author} {\bibfnamefont {A.~W.}\ \bibnamefont
  {Chao}},\ }\href@noop {} {\emph {\bibinfo {title} {{Physics of collective
  beam instabilities in high-energy accelerators}}}}\ (\bibinfo  {publisher}
  {John Wiley \& Sons, Inc.},\ \bibinfo {year} {1993})\BibitemShut {NoStop}%
\bibitem [{\citenamefont {Inc.}(2021)}]{Mathematica}%
  \BibitemOpen
  \bibfield  {author} {\bibinfo {author} {\bibfnamefont {W.~R.}\ \bibnamefont
  {Inc.}},\ }\href {https://www.wolfram.com/mathematica} {\enquote {\bibinfo
  {title} {Mathematica, {V}ersion 12.1},}\ } (\bibinfo {year} {2021}),\
  \bibinfo {note} {champaign, IL, 2021}\BibitemShut {NoStop}%
\bibitem [{\citenamefont {Sagan}(2006)}]{Sagan:Bmad2006}%
  \BibitemOpen
  \bibfield  {author} {\bibinfo {author} {\bibfnamefont {D.}~\bibnamefont
  {Sagan}},\ }\bibfield  {booktitle} {\emph {\bibinfo {booktitle}
  {{Computational accelerator physics. Proceedings, 8th International
  Conference, ICAP 2004, St. Petersburg, Russia, June 29-July 2, 2004}}},\
  }\href {\doibase https://doi.org/10.1016/j.nima.2005.11.001} {\bibfield
  {journal} {\bibinfo  {journal} {Nucl. Instrum. Meth.}\ }\textbf {\bibinfo
  {volume} {A558}},\ \bibinfo {pages} {356} (\bibinfo {year} {2006})},\
  \bibinfo {note} {proceedings of the 8th International Computational
  Accelerator Physics Conference}\BibitemShut {NoStop}%
%%CITATION = NUIMA,A558,356;%%
\bibitem [{\citenamefont
  {Leemann}(2012{\natexlab{b}})}]{PhysRevSTAB.15.050705}%
  \BibitemOpen
  \bibfield  {author} {\bibinfo {author} {\bibfnamefont {S.~C.}\ \bibnamefont
  {Leemann}},\ }\href {\doibase 10.1103/PhysRevSTAB.15.050705} {\bibfield
  {journal} {\bibinfo  {journal} {Phys. Rev. ST Accel. Beams}\ }\textbf
  {\bibinfo {volume} {15}},\ \bibinfo {pages} {050705} (\bibinfo {year}
  {2012}{\natexlab{b}})}\BibitemShut {NoStop}%
\bibitem [{\citenamefont {Rast}\ \emph {et~al.}(2011)\citenamefont {Rast},
  \citenamefont {Atkinson}, \citenamefont {Dirsat}, \citenamefont {Dressler},\
  and\ \citenamefont {Kuske}}]{Rast:2011zz}%
  \BibitemOpen
  \bibfield  {author} {\bibinfo {author} {\bibfnamefont {H.}~\bibnamefont
  {Rast}}, \bibinfo {author} {\bibfnamefont {T.}~\bibnamefont {Atkinson}},
  \bibinfo {author} {\bibfnamefont {M.}~\bibnamefont {Dirsat}}, \bibinfo
  {author} {\bibfnamefont {O.}~\bibnamefont {Dressler}}, \ and\ \bibinfo
  {author} {\bibfnamefont {P.}~\bibnamefont {Kuske}},\ }\href@noop {}
  {\bibfield  {journal} {\bibinfo  {journal} {Conf. Proc. C}\ }\textbf
  {\bibinfo {volume} {110904}},\ \bibinfo {pages} {3396} (\bibinfo {year}
  {2011})}\BibitemShut {NoStop}%
\bibitem [{\citenamefont {Liu}\ \emph {et~al.}(2016)\citenamefont {Liu},
  \citenamefont {Resende}, \citenamefont {Rodrigues},\ and\ \citenamefont
  {de~S\'a}}]{Liu:2016rcc}%
  \BibitemOpen
  \bibfield  {author} {\bibinfo {author} {\bibfnamefont {L.}~\bibnamefont
  {Liu}}, \bibinfo {author} {\bibfnamefont {X.}~\bibnamefont {Resende}},
  \bibinfo {author} {\bibfnamefont {A.}~\bibnamefont {Rodrigues}}, \ and\
  \bibinfo {author} {\bibfnamefont {F.}~\bibnamefont {de~S\'a}},\ }in\ \href
  {\doibase 10.18429/JACoW-IPAC2016-THPMR011} {\emph {\bibinfo {booktitle}
  {{7th International Particle Accelerator Conference}}}}\ (\bibinfo {year}
  {2016})\BibitemShut {NoStop}%
\bibitem [{\citenamefont {Alexandre}\ \emph {et~al.}(2021)\citenamefont
  {Alexandre}, \citenamefont {Fekih}, \citenamefont {Letrésor}, \citenamefont
  {Thoraud}, \citenamefont {{da Silva Castro}}, \citenamefont {Bouvet},
  \citenamefont {Breunlin}, \citenamefont {Åke Andersson},\ and\ \citenamefont
  {{Fernandes Tavares}}}]{ALEXANDRE2021164739}%
  \BibitemOpen
  \bibfield  {author} {\bibinfo {author} {\bibfnamefont {P.}~\bibnamefont
  {Alexandre}}, \bibinfo {author} {\bibfnamefont {R.~B.~E.}\ \bibnamefont
  {Fekih}}, \bibinfo {author} {\bibfnamefont {A.}~\bibnamefont {Letrésor}},
  \bibinfo {author} {\bibfnamefont {S.}~\bibnamefont {Thoraud}}, \bibinfo
  {author} {\bibfnamefont {J.}~\bibnamefont {{da Silva Castro}}}, \bibinfo
  {author} {\bibfnamefont {F.}~\bibnamefont {Bouvet}}, \bibinfo {author}
  {\bibfnamefont {J.}~\bibnamefont {Breunlin}}, \bibinfo {author} {\bibnamefont
  {Åke Andersson}}, \ and\ \bibinfo {author} {\bibfnamefont {P.}~\bibnamefont
  {{Fernandes Tavares}}},\ }\href {\doibase
  https://doi.org/10.1016/j.nima.2020.164739} {\bibfield  {journal} {\bibinfo
  {journal} {Nuclear Instruments and Methods in Physics Research Section A:
  Accelerators, Spectrometers, Detectors and Associated Equipment}\ }\textbf
  {\bibinfo {volume} {986}},\ \bibinfo {pages} {164739} (\bibinfo {year}
  {2021})}\BibitemShut {NoStop}%
\bibitem [{\citenamefont {Sun}\ \emph {et~al.}(2020)\citenamefont {Sun},
  \citenamefont {Amstutz}, \citenamefont {Hellert}, \citenamefont {Leemann},
  \citenamefont {Steier}, \citenamefont {Swenson},\ and\ \citenamefont
  {Venturini}}]{PhysRevAccelBeams.23.010702}%
  \BibitemOpen
  \bibfield  {author} {\bibinfo {author} {\bibfnamefont {C.}~\bibnamefont
  {Sun}}, \bibinfo {author} {\bibfnamefont {P.}~\bibnamefont {Amstutz}},
  \bibinfo {author} {\bibfnamefont {T.}~\bibnamefont {Hellert}}, \bibinfo
  {author} {\bibfnamefont {S.~C.}\ \bibnamefont {Leemann}}, \bibinfo {author}
  {\bibfnamefont {C.}~\bibnamefont {Steier}}, \bibinfo {author} {\bibfnamefont
  {C.}~\bibnamefont {Swenson}}, \ and\ \bibinfo {author} {\bibfnamefont
  {M.}~\bibnamefont {Venturini}},\ }\href {\doibase
  10.1103/PhysRevAccelBeams.23.010702} {\bibfield  {journal} {\bibinfo
  {journal} {Phys. Rev. Accel. Beams}\ }\textbf {\bibinfo {volume} {23}},\
  \bibinfo {pages} {010702} (\bibinfo {year} {2020})}\BibitemShut {NoStop}%
\end{thebibliography}%

\end{document}